%% file: main.tex
\documentclass[conference]{IEEEtran}
\let\labelindent\relax

\usepackage{subcaption}
\usepackage{times}
\usepackage{array,multirow}
\usepackage{xspace}
\usepackage{xcolor}
\usepackage{color, colortbl}
\usepackage{balance}
\usepackage{paralist}
\usepackage{amsmath}
\usepackage{amsfonts,amssymb,amsthm}
\usepackage{graphicx}
\usepackage{pifont}
\usepackage{array}
\usepackage{booktabs}
\usepackage{tikz}
\usepackage{bm}
\usepackage{enumitem}
\usepackage{url}
\usepackage[draft]{todonotes}
\usepackage{letltxmacro}
\usepackage{cleveref}

\input{macros}

\newtheorem{theorem}{Theorem}
\usepackage{mathtools} 

\ifCLASSINFOpdf
\else
\fi
\begin{document}
%
\title{\sysname : Threshold Issuance Selective Disclosure Credentials with Applications to Distributed Ledgers}

\author{
\IEEEauthorblockN{Alberto Sonnino\IEEEauthorrefmark{1}\IEEEauthorrefmark{2}, 
Mustafa Al-Bassam\IEEEauthorrefmark{1}\IEEEauthorrefmark{2}, 
Shehar Bano\IEEEauthorrefmark{1}\IEEEauthorrefmark{2},
Sarah Meiklejohn\IEEEauthorrefmark{1} 
and George Danezis\IEEEauthorrefmark{1}\IEEEauthorrefmark{2}}
\IEEEauthorblockA{\IEEEauthorrefmark{1} University College London, United Kingdom} 
\IEEEauthorblockA{\IEEEauthorrefmark{2} \texttt{chainspace.io}}
}


%


\IEEEoverridecommandlockouts
\makeatletter\def\@IEEEpubidpullup{6.5\baselineskip}\makeatother
\IEEEpubid{\parbox{\columnwidth}{
    Network and Distributed Systems Security (NDSS) Symposium 2019\\
    24-27 February 2019, San Diego, CA, USA\\
    ISBN 1-891562-55-X\\
    https://dx.doi.org/10.14722/ndss.2019.23272\\
    www.ndss-symposium.org
}
\hspace{\columnsep}\makebox[\columnwidth]{}}

\maketitle

\begin{abstract}
\sysname is a novel selective disclosure credential scheme supporting distributed threshold issuance, public and private attributes, re-randomization, and multiple unlinkable selective attribute revelations. \sysname integrates with 
\blockchains to ensure confidentiality, authenticity and availability even when a subset of credential issuing authorities are malicious or offline. We implement and evaluate a generic \sysname smart contract library for \chainspace and \ethereum; and present three applications related to anonymous payments, electronic petitions, and distribution of proxies for censorship resistance.
\sysname uses short and computationally efficient credentials, and our evaluation shows that most \sysname cryptographic primitives take just a few milliseconds on average, with verification taking the longest time (10 milliseconds). 
\end{abstract}


%

\input{sections/introduction}
\input{sections/architecture}
\input{sections/construction}

\input{sections/implementation}

\input{sections/applications}
\input{sections/evaluation}

\input{sections/related}

\input{sections/limitations}
\input{sections/conclusion}
\input{acknowledgements}

\bibliographystyle{IEEEtranS}
\bibliography{references}

\appendices
\input{appendices/security}

\input{appendices/multi_attribute_credentials}

\input{appendices/eth-tumbler}




%
%
%
%

\end{document}

%% file: macros.tex
\newcommand\sysname{Coconut\xspace}

\newcommand\chainspace{Chainspace\xspace}
\newcommand\ethereum{Ethereum\xspace}
\newcommand\hyperledger{Hyperledger\xspace}
\newcommand\omniledger{Omniledger\xspace}
\newcommand\elgamal{El-Gamal\xspace}

\newcommand\algorithm[1]{\textsf{ #1}}

\newcommand\hashtopoint{H\xspace}

\newcommand\definition[2]{\ding{118}\xspace \textsf{#1}\xspace$\bm{\rightarrow}$\xspace(#2):\xspace}

\newcommand\blockchain{blockchain\xspace}
\newcommand\blockchains{blockchains\xspace}

\newcommand{\etal}{\textit{et al.}\@\xspace}

\newcommand{\ie}{\textit{i.e.,}\@\xspace}
\newcommand{\via}{\textit{via}\@\xspace}

\def\first{({\it i})\xspace}
\def\second{({\it ii})\xspace}
\def\third{({\it iii})\xspace}
\def\fourth{({\it iv})\xspace}

\newcommand{\yes}{\ding{51}}
\newcommand{\no}{\ding{55}}

\DeclareRobustCommand\pie[1]{%
    \tikz[every node/.style={inner sep=0,outer sep=0, scale=1.5}]{
        \node[minimum size=1.5ex] at (0,-1.5ex) {}; 
	    \draw[fill=white] (0,-1.5ex) circle (0.75ex); \draw[fill=black] (0.75ex,-1.5ex) arc (0:#1:0.75ex); 
	}%
}
\def\L{\pie{0}}	
\def\M{\pie{-180}}	
\def\H{\pie{360}}	


%% file: sections/introduction.tex
\section{Introduction} \label{introduction}

Selective disclosure credentials~\cite{cl,amac} allow the issuance of a credential to a user, and the subsequent unlinkable revelation (or `showing') of some of the attributes it encodes to a verifier for the purposes of authentication, authorization or to implement electronic cash. However, established schemes have shortcomings. Some entrust a single issuer with the credential signature key, allowing a malicious issuer to forge any credential or electronic coin. Other schemes do not provide the necessary efficiency, re-randomization, or blind issuance properties necessary to implement practical selective disclosure credentials. No existing scheme provides all of efficiency, threshold distributed issuance, private attributes, re-randomization, and unlinkable multi-show selective disclosure.  

The lack of efficient general purpose selective disclosure credentials impacts platforms that support `smart contracts', 
such as \ethereum~\cite{ethereum}, \hyperledger~\cite{hyperledger} and \chainspace~\cite{chainspace}. 
They all share the limitation that verifiable smart contracts may only perform operations recorded on a public \blockchain.
Moreover, the security models of these systems generally assume that integrity should hold in the presence of a threshold number of dishonest or faulty nodes (Byzantine fault tolerance); it is desirable for similar assumptions to hold for multiple credential issuers (threshold issuance).

Issuing credentials through smart contracts would be very desirable: a smart contract could conditionally issue user credentials depending on the state of the \blockchain, or attest some claim about a user operating through the contract---such as their identity, attributes, or even the balance of their wallet. 
This is not possible, as current selective credential schemes would either entrust a single party as an issuer, or would not provide appropriate efficiency, re-randomization, blind issuance and selective disclosure capabilities (as in the case of threshold signatures~\cite{back2014enabling}). For example, the \hyperledger system supports CL credentials~\cite{cl} through a trusted third party issuer, illustrating their usefulness, but also their fragility against the issuer becoming malicious. Garman~\etal~\cite{dac} present a decentralized anonymous credentials system integrated into distributed ledgers; they provide the ability to issue publicly verifiable claims without central issuers, but do not focus on threshold issuance or on general purpose credentials, and showing credentials requires expensive double discrete-logarithm proofs.

\sysname addresses these challenges, and allows a subset of decentralized mutually distrusting authorities to jointly issue credentials, on public or private attributes. Those credentials cannot be forged by users, or any small subset of potentially corrupt authorities. Credentials can be re-randomized before selected attributes are shown to a verifier, protecting privacy even in the case in which all authorities and verifiers collude. The \sysname scheme is based on a threshold issuance signature scheme that allows partial claims to be aggregated into a single credential. Mapped to the context of permissioned and semi-permissioned \blockchains, \sysname allows collections of authorities in charge of maintaining a \blockchain, or a side chain~\cite{back2014enabling} based on a federated peg, 
to jointly issue selective disclosure credentials.

\sysname uses short and computationally efficient credentials, and efficient revelation of selected attributes and verification protocols.
Each partial credential and the consolidated credential is composed of exactly two group elements. The size of the credential remains constant regardless of the number of attributes or authorities/issuers. Furthermore, after a one-time setup phase where the users collect and aggregate a threshold number of verification keys from the authorities, the attribute showing and verification are $O(1)$ in terms of both cryptographic computations and communication of cryptographic material---irrespective of the number of authorities. Our evaluation of the \sysname primitives shows very promising results. Verification takes about 10ms, while signing a private attribute is about 3 times faster.
 The latency is about 600 ms when the client aggregates partial credentials from 10 authorities distributed across the world.



\noindent\textbf{Contribution.} This paper makes three key contributions:

\begin{itemize}
\setlength\itemsep{0em}
    \item We describe the signature schemes underlying \sysname, including how
key generation, distributed issuance, aggregation and verification of
signatures operate (\Cref{architecture,construction}). The scheme is an
extension and hybrid of the Waters signature scheme~\cite{waters}, the BGLS
signature~\cite{bgls}, and the signature scheme of
Pointcheval and Sanders~\cite{pointcheval}. This is the first general purpose, fully distributed threshold issuance, re-randomizable, multi-show credential scheme of which we are aware.
    
    \item We use \sysname to implement a generic smart contract library for \chainspace~\cite{chainspace} and one for \ethereum~\cite{ethereum}, performing public and private attribute issuance, aggregation, randomization and selective disclosure (\Cref{implementation}). We evaluate their performance and cost within those platforms (\Cref{evaluation}).
    
    \item We design three applications using the \sysname contract library: a coin tumbler providing payment anonymity; a privacy preserving electronic petitions; and a proxy distribution system for a censorship resistance system (\Cref{applications}). We implement and evaluate the first two applications on the \chainspace platform, and provide a security and performance evaluation (\Cref{evaluation}).

\end{itemize}


%% file: sections/architecture.tex
\section{Overview of \sysname} 

\label{architecture}

\begin{figure}[t]
    \centering
    \includegraphics[width=.35\textwidth]{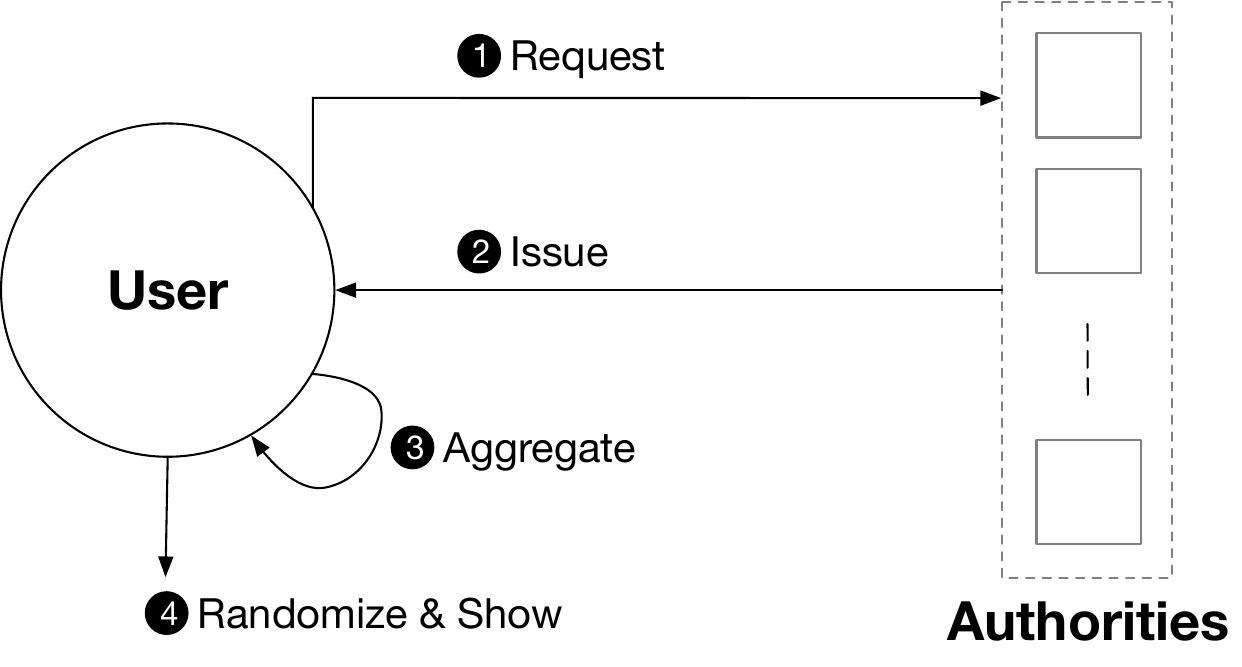}
    \caption{\footnotesize A high-level overview of \sysname architecture.}
    \label{fig:gen_arch}
\end{figure}


\sysname is a selective disclosure credential system, supporting threshold credential issuance of public and private attributes, re-randomization of credentials to support multiple unlinkable revelations, and the ability to selectively disclose a subset of attributes. It is embedded into a smart contract library that can be called from other contracts to issue credentials.

The \sysname architecture is illustrated in \Cref{fig:gen_arch}. Any \sysname user may send a \sysname \emph{request} command to a set of \sysname signing authorities; this command specifies a set of public or encrypted private attributes to be certified into the credential~(\ding{202}). Then, each authority answers with an \emph{issue} command delivering a partial credential~(\ding{203}). Any user can collect a threshold number of shares, aggregate them to form a single consolidated credential, and re-randomize it (\ding{204}). The use of the credential for authentication is however restricted to a user who knows the private attributes embedded in the credential---such as a private key.  The user who owns the credentials can then execute the \emph{show} protocol to selectively disclose attributes or statements about them~(\ding{205}). The showing protocol is publicly verifiable, and may be publicly recorded. \sysname has the following design goals:

\begin{itemize}[leftmargin=2em, labelindent=0em]
\setlength\itemsep{0.5em}

\item \textbf{Threshold authorities:} Only a subset of the authorities is required to issue partial credentials in order to allow the users to generate a consolidated credential~\cite{boldyreva2002efficient}. The communication complexity of the \emph{request} and \emph{issue} protocol is thus $O(t)$, where $t$ is the size of the subset of authorities. Furthermore, it is impossible to generate a consolidated credential from fewer than $t$ partial credentials.
\item \textbf{Blind issuance \& Unlinkability:} The authorities issue the credential without learning any additional information about the private attributes embedded in the credential. Furthermore, it is impossible to link multiple showings of the credentials with each other, or the issuing transcript, even if all the authorities collude (see \Cref{definitions}).
\item \textbf{Non-interactivity:} The authorities may operate independently of each other, following a simple key distribution and setup phase to agree on public security and cryptographic parameters---they do not need to synchronize or further coordinate their activities.

\item \textbf{Liveness:} \sysname guarantees liveness as long as a threshold number of authorities remains honest and weak synchrony assumptions holds for the key distribution~\cite{cryptoeprint:2012:377}.

\item \textbf{Efficiency:} The credentials and all zero-knowledge proofs involved in the protocols are short and computationally efficient. After aggregation and re-randomization, the attribute showing and verification involve only a single consolidated credential, and are therefore $O(1)$ in terms of both cryptographic computations and communication of cryptographic material---no matter the number of authorities.
\item \textbf{Short credentials:} Each partial credential---as well as the consolidated credential---is composed of exactly two group elements, no matter the number of authorities or the number of attributes embedded in the credentials.
\end{itemize}

As a result, a large number of authorities may be used to issue credentials, without significantly affecting efficiency.

%% file: sections/construction.tex
\section{The \sysname Construction} \label{construction}
We introduce the cryptographic primitives supporting the \sysname architecture, step by step from the design of Pointcheval and Sanders~\cite{pointcheval} and Boneh~\etal~\cite{bls, bgls} to the full \sysname scheme.

\begin{itemize}[leftmargin=2em, labelindent=0em]
\setlength\itemsep{0.5em}

\item \textbf{Step 1:} We first recall (\Cref{pointcheval_recall}) the scheme of Pointcheval~\etal~\cite{pointcheval} for single-attribute credentials. We present its limitations preventing it from meeting our design goals presented in \Cref{architecture}, and we show how to incorporate principles from Boneh~\etal~\cite{bls} to overcome them. 

\item \textbf{Step 2:} We introduce (\Cref{threshold_credentials_scheme}) the \emph{\sysname threshold credentials scheme}, which has all the properties of Pointcheval and Sanders~\cite{pointcheval} and Boneh~\etal~\cite{bls}, and allows us to achieve all our design goals.

\item \textbf{Step 3:} Finally, we extend (\Cref{multi_message_scheme}) our schemes to support credentials embedding $q$ distinct attributes $(m_1,\dots,m_{q})$ simultaneously.
\end{itemize}

\subsection{Notations and Assumptions} \label{background_and_assumptions}
We present the notation used in the rest of the paper, as well as the security assumptions on which our primitives rely.

\paragraph{Zero-knowledge proofs} 
Our credential scheme uses non-interactive zero-knowledge proofs to assert knowledge and relations over discrete logarithm values. We represent these non-interactive zero-knowledge proofs with the notation introduced by Camenisch~\etal~\cite{camenisch1997proof}:
\begin{equation}\nonumber
{\rm NIZK}\{(x,y,\dots): \textrm{statements about } x, y, \dots\}
\end{equation}
which denotes proving in zero-knowledge that the secret values $(x,y,\dots)$ (all other values are public) satisfy the statements after the colon. 


\paragraph{Cryptographic assumptions}
\sysname requires groups $(\mathbb{G}_1,\mathbb{G}_2,\mathbb{G}_T)$ of prime order $p$ with a bilinear map $e:\mathbb{G}_1 \times \mathbb{G}_2 \rightarrow \mathbb{G}_T$ and satisfying the following properties: \first\emph{Bilinearity} means that for all $g_1 \in \mathbb{G}_1$, $g_2 \in \mathbb{G}_2$ and $(a,b) \in \mathbb{F}_p^2$, \; $e(g_1^a,g_2^b) = e(g_1,g_2)^{ab}$; \second\emph{Non-degeneracy} means that for all $g_1 \in \mathbb{G}_1$, $g_2 \in \mathbb{G}_2$, $e(g_1,g_2) \neq 1$; \third\emph{Efficiency} implies the map $e$ is efficiently computable; \fourth furthermore, $\mathbb{G}_1 \neq \mathbb{G}_2$, and there is no efficient homomorphism between $\mathbb{G}_1$ and $\mathbb{G}_2$.
The type-3 pairings are efficient~\cite{galbraith2008pairings}. They support the XDH assumption which implies the difficulty of the Computational co-Diffie-Hellman (co-CDH) problem in $\mathbb{G}_1$ and $\mathbb{G}_2$, and the difficulty of the Decisional Diffie-Hellman (DDH) problem in $\mathbb{G}_1$~\cite{bls}.

\sysname also relies on a cryptographically secure hash function $\hashtopoint$, hashing an element $\mathbb{G}_1$ into an other element of $\mathbb{G}_1$, namely $\hashtopoint: \mathbb{G}_1\rightarrow\mathbb{G}_1 $. We implement this function by serializing the $(x,y)$ coordinates of the input point and applying a full-domain hash function to hash this string into an element of $\mathbb{G}_1$ (as Boneh~\etal~\cite{bls}).


\paragraph{Threshold and communication assumptions}
\sysname assumes honest majority ($n/2 < t$) to prevent malicious authorities from issuing credentials arbitrarily. \sysname authorities do not need to communicate with each other; users wait for $t$-out-of-$n$ replies (in any order of arrival) and aggregate them into a consolidated credential; thus \sysname implicitly assumes an asynchronous setting. However, our current implementations rely on the distributed key generation protocol of Kate~\etal~\cite{cryptoeprint:2012:377}, which requires \first weak synchrony for liveness (but not for safety), and \second at most one third of dishonest authorities.

\subsection{Scheme Definitions and Security Properties}\label{definitions}

We present the protocols that comprise a threshold credentials scheme:
\begin{description}[leftmargin=1em, labelindent=0em]
\setlength\itemsep{0.5em}

\item[\definition{Setup($1^\lambda$)}{$params$}] defines the system parameters $params$ with respect to the security parameter $\lambda$. These parameters are publicly available.

\item[\definition{KeyGen($params$)}{$sk,vk$}] is run by the authorities to generate their secret key $sk$ and verification key $vk$ from the public $params$.

\item[\definition{AggKey($vk_1, \dots, vk_t$)}{$vk$}] is run by whoever wants to verify a credential to aggregate any subset of $t$ verification keys $vk_i$ into a single consolidated verification key $vk$. \algorithm{AggKey} needs to be run only once.

\item[\definition{IssueCred($m,\phi$)}{$\sigma$}] is an interactive protocol between a user and each authority, by which the user obtains a credential $\sigma$ embedding the private attribute $m$ satisfying the statement $\phi$. 

\item[\definition{AggCred($\sigma_1, \dots, \sigma_t$)}{$\sigma$}] is run by the user to aggregate any subset of $t$ partial credentials $\sigma_i$ into a single consolidated credential.

\item[\definition{ProveCred($vk, m, \phi'$)}{$\Theta,\phi'$}] is run by the user to compute a proof $\Theta$ of possession of a credential certifying that the private attribute $m$ satisfies the statement $\phi'$ (under the corresponding verification key $vk$).

\item[\definition{VerifyCred($vk, \Theta, \phi'$)}{$true/false$}] is run by whoever wants to verify a credential embedding a private attribute satisfying the statement $\phi'$, using the verification key $vk$ and cryptographic material $\Theta$ generated by \textsf{ProveCred}. 
\end{description}


A threshold credential scheme must satisfy the following security properties:
\begin{description}[leftmargin=1em, labelindent=0em]
\setlength\itemsep{0.5em}
\item [Unforgeability: ] It must be unfeasible for an adversarial user to convince an honest verifier that they are in possession of a credential if they are in fact not (i.e., if they have not received valid partial credentials from at least $t$ authorities).

\item [Blindness: ] It must be unfeasible for an adversarial authority to learn any information about the attribute $m$ during the execution of the \algorithm{IssueCred} protocol, except for the fact that $m$ satisfies $\phi$.

\item [Unlinkability / Zero-knowledge: ] It must be unfeasible for an adversarial verifier (potentially working with an adversarial authority) to learn anything about the attribute $m$, except that it satisfies $\phi'$, or to link the execution of \algorithm{ProveCred} with either another execution of \algorithm{ProveCred} or with the execution of \algorithm{IssueCred} (for a given attribute $m$).
\end{description}

\subsection{Foundations of Coconut}\label{pointcheval_recall}
Before giving the full \sysname construction, we first recall the credentials scheme proposed by Pointcheval and Sanders~\cite{pointcheval}; their construction has the same properties as CL-signatures~\cite{cl} but is more efficient. The scheme works in a bilinear group $(\mathbb{G}_1,\mathbb{G}_2,\mathbb{G}_T)$ of type 3, with a bilinear map $e:\mathbb{G}_1 \times \mathbb{G}_2 \rightarrow \mathbb{G}_T$ as described in \Cref{background_and_assumptions}. 

\begin{description}[leftmargin=1em, labelindent=0em]
\setlength\itemsep{0.5em}
\item[\definition{P.Setup($1^\lambda$)}{$params$}] Choose a bilinear group $(\mathbb{G}_1,\mathbb{G}_2,\mathbb{G}_T)$ with order $p$, where $p$ is a $\lambda$-bit prime number. Let $g_1$ be a generator of $\mathbb{G}_1$, and $g_2$ a generator of $\mathbb{G}_2$. The system parameters are $params=(\mathbb{G}_1, \mathbb{G}_2, \mathbb{G}_T, p, g_1, g_2)$. 

\item[\definition{P.KeyGen($params$)}{$sk,vk$}] Choose a random secret key $sk = (x,y) \in \mathbb{F}_p^2$. Parse $params=(\mathbb{G}_1, \mathbb{G}_2, \mathbb{G}_T, p, g_1, g_2)$, and publish the verification key $vk = (g_2, \alpha,\beta) = (g_2, g_2^x,g_2^y)$.

\item[\definition{P.Sign($params, sk, m$)}{$\sigma$}] Parse $sk = (x, y)$. Pick a random $r \in \mathbb{F}_p$ and set $h=g_1^{r}$. Output $\sigma = (h, s) = (h, h^{x+y\cdot m})$.

\item[\definition{P.Verify($params, vk, m, \sigma$)}{$true/false$}] Parse $vk = (g_2, \alpha,\beta)$ and $\sigma = (h, s)$. Output $true$ if $h\neq1$ and $e(h,\alpha\beta^m)=e(s,g_2)$; otherwise output $false$.

\end{description}

The signature $\sigma=(h,s)$ is randomizable by choosing a random $r' \in \mathbb{F}_p$ and computing $\sigma'=(h^{r'},s^{r'})$. The above scheme can be modified to obtain credentials on a private attribute: to run \algorithm{IssueCred} the user first picks a random $t \in \mathbb{F}_p$, computes the commitment $c_p=g_1^tY^m$ to the message $m$, where $Y=g_1^y$; and sends it to a single authority along with a zero-knowledge proof of the opening of the commitment. The authority verifies the proof, picks a random $u \in \mathbb{F}_p$, and returns $\widetilde{\sigma}=(h,\widetilde{s})=(g^u,(Xc_p)^u)$ where $X=g_1^x$. The user unblinds the signature by computing $\sigma=(h,\widetilde{s}(h)^{-t})$, and this value acts as the credential.

This scheme provides blindness, unlinkability, efficiency and short credentials; but it does not support threshold issuance and therefore does not achieve our design goals. This limitation comes from the \textsf{P.Sign} algorithm---the issuing authority computes the credentials using a private and self-generated random number $r$ which prevents the scheme from being efficiently distributed to a multi-authority setting\footnote{The original paper of Pointcheval and Sanders~\cite{pointcheval} proposes a sequential aggregate signature protocol that is  unsuitable for threshold credentials issuance (see \Cref{related}).}. To overcome that limitation, we take advantage of a concept introduced by BLS signatures~\cite{bls}; exploiting a hash function $\hashtopoint: \mathbb{F}_p\rightarrow\mathbb{G}_1$ to compute the group  element $h=\hashtopoint(m)$. The next section describes how \sysname incorporates these concepts to achieve all our design goals.

\subsection{The \sysname Threshold Credential Scheme} \label{threshold_credentials_scheme}
We introduce the \emph{\sysname} threshold credential scheme, allowing users to obtain a partial credential $\sigma_i$ on a private or public attribute $m$. In a system with $n$ authorities, a $t$-out-of-$n$ threshold credentials scheme offers great flexibility as the users need to collect only $n/2< t \leq n$ of these partial credentials in order to recompute the consolidated credential (both $t$ and $n$ are scheme parameters). 


\paragraph{Cryptographic primitives} For the sake of simplicity, we describe below a key generation algorithm \algorithm{TTPKeyGen} as executed by a trusted third party; this protocol can however be executed in a distributed way as illustrated by Gennaro~\etal~\cite{gennaro1999secure} under a synchrony assumption, and as illustrated by Kate~\etal~\cite{cryptoeprint:2012:377} under a weak synchrony assumption. Adding and removing authorities implies a re-run of the key generation algorithm---this limitation is inherited from the underlying Shamir's secret sharing protocol~\cite{shamir1979share} and can be mitigated using techniques introduced by Herzberg~\etal~\cite{herzberg1995proactive}.

\begin{description}[leftmargin=1em, labelindent=0em]
\setlength\itemsep{0.5em}
\item[\definition{Setup($1^\lambda$)}{$params$}] Choose a bilinear group $(\mathbb{G}_1,\mathbb{G}_2,\mathbb{G}_T)$ with order $p$, where $p$ is a $\lambda$-bit prime number. Let $g_1, h_1$ be generators of $\mathbb{G}_1$, and $g_2$ a generator of $\mathbb{G}_2$. The system parameters are $params=(\mathbb{G}_1, \mathbb{G}_2, \mathbb{G}_T, p, g_1, g_2, h_1)$. 

\item[\definition{TTPKeyGen($params, t, n$)}{$sk,vk$}] Pick\footnote{This algorithm can be turned into the \textsf{KeyGen} and \textsf{AggKey} algorithms described in \Cref{definitions} using techniques illustrated by Gennaro~\etal~\cite{gennaro1999secure} or Kate~\etal~\cite{cryptoeprint:2012:377}.} two polynomials $v,w$ of degree $t-1$ with coefficients in $\mathbb{F}_p$, and set $(x,y) = (v(0), w(0))$. Issue to each authority $i \in [1, \dots, n]$ a secret key $sk_i = (x_i,y_i) = (v(i), w(i))$, and publish their verification key $vk_i$ = $(g_2,\alpha_i,\beta_i) = (g_2,g_2^{x_i},g_2^{y_i})$.

\item[\definition{IssueCred($m, \phi$)}{$\sigma$}] Credentials issuance is composed of three algorithms:
\begin{description}[leftmargin=1em, labelindent=0em]
\setlength\itemsep{0.5em}
\item \definition{PrepareBlindSign($m, \phi$)}{$d,\Lambda,\phi$} The users generate an \elgamal key-pair $(d, \gamma=g_1^{d})$; pick a random $o\in\mathbb{F}_p$,  compute the commitment $c_m$ and the group element $h\in\mathbb{G}_1$ as follows:
\begin{equation}\nonumber
c_m = g_1^m h_1^o \qquad{\rm and}\qquad h = \hashtopoint(c_m)
\end{equation} 
Pick a random $k \in \mathbb{F}_p$ and compute an \elgamal encryption of $m$ as below:
\begin{equation}\nonumber
c = Enc(h^m)=(g_1^k,\gamma^k h^m)
\end{equation}
Output $(d, \Lambda=(\gamma, c_m, c, \pi_{s}), \phi)$, where $\phi$ is an application-specific predicate satisfied by $m$, and $\pi_{s}$ is defined by:
\begin{eqnarray}\nonumber
\pi_{s} &=& {\rm NIZK}\{(d, m, o, k): \gamma = g_1^d \;\land\; c_m=g_1^mh_1^o\\ \nonumber
 && \land\; c = (g_1^k,\gamma^k h^m) \;\land\;  \phi(m)=1\}
 \end{eqnarray}

\item \definition{BlindSign($sk_i, \Lambda, \phi$)}{$\tilde{\sigma}_i$} The authority $i$ parses $\Lambda=(\gamma, c_m, c, \pi_{s})$, $sk_i=(x_i,y_i)$, and $c=(a,b)$. Recompute $h = \hashtopoint(c_m)$. Verify the proof  $\pi_{s}$ using $\gamma$, $c_m$ and $\phi$; if the proof is valid, build $\tilde{c}_i=(a^y,h^{x_i}b^{y_i})$ and output $\tilde{\sigma}_i = (h, \tilde{c}_i)$; otherwise output $\perp$ and stop the protocol.

\item \definition{Unblind($\tilde{\sigma}_i, d$)}{$\sigma_i$} The users parse $\tilde{\sigma}_i=(h, \tilde{c})$ and $\tilde{c}=(\tilde{a},\tilde{b})$; compute $\sigma_i = (h,\tilde{b}(\tilde{a})^{-d})$. Output $\sigma_i$.
 \end{description}
 
\item[\definition{AggCred($\sigma_1, \dots, \sigma_t$)}{$\sigma$}] Parse each $\sigma_i$ as $(h,s_i)$ for $i \in [1, \dots, t]$. Output $(h,\prod^t_{i=1} s_i^{l_i})$, where $l$ is the Lagrange coefficient:
\begin{equation}\nonumber
l_i = \left[\prod^t_{j=1, j\neq i} (0-j)\right] \left[\prod^t_{j=1, j\neq i} (i-j)\right]^{-1} \;{\rm mod}\; p
\end{equation}

\item[\definition{ProveCred($vk, m, \sigma, \phi'$)}{$\Theta,\phi'$}] Parse $\sigma=(h,s)$ and $vk=(g_2,\alpha,\beta)$. Pick at random $r',r \in \mathbb{F}_p^2$; set $\sigma'=(h',s')=(h^{r'},s^{r'})$; build $\kappa = \alpha\beta^m g_2^r$ and $\nu=\left(h'\right)^r$. Output $(\Theta=(\kappa, \nu, \sigma',\pi_v),\phi')$, where $\phi'$ is an application-specific predicate satisfied by $m$, and $\pi_v$ is:
\begin{equation}\nonumber
    \pi_v={\rm NIZK}\{(m,r): \kappa=\alpha\beta^m g_2^r \ \land \ \nu=\left(h'\right)^r \ \land \  \phi'(m)=1\} 
\end{equation}

\item[\definition{VerifyCred($vk, \Theta, \phi'$)}{$true/false$}] Parse $\Theta = (\kappa, \nu, \sigma',\pi_v)$ and $\sigma'=(h',s')$; verify $\pi_v$ using $vk$ and $\phi'$. Output $true$ if the proof verifies, $h'\neq1$ and $e(h',\kappa)=e(s'\nu,g_2)$; otherwise output $false$.
\end{description}

\begin{figure}[t]
    \centering
    \includegraphics[width=.48\textwidth]{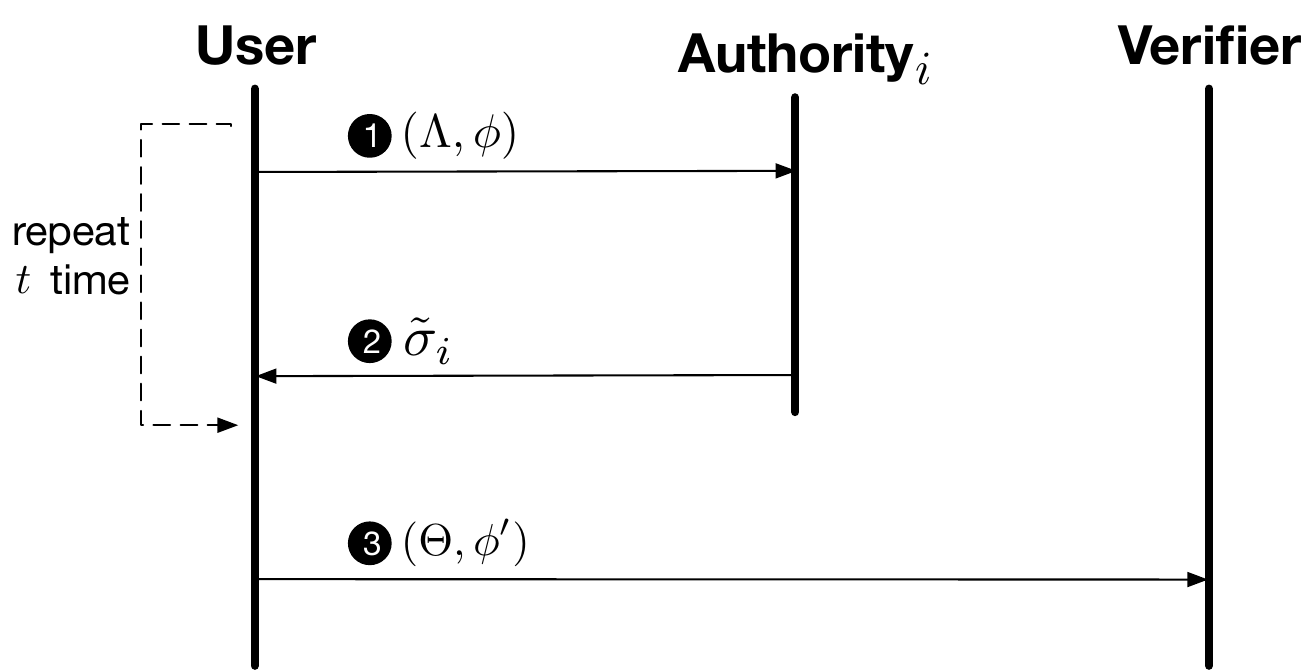}
    \caption{
    \footnotesize \sysname threshold credentials protocol exchanges. 
    }
    \label{fig:protocol_priv}
\end{figure}

\paragraph{Correctness and explanation} The \algorithm{Setup} algorithm generates the public parameters. Credentials are elements of $\mathbb{G}_1$, while verification keys are elements of $\mathbb{G}_2$. \Cref{fig:protocol_priv} illustrates the protocol exchanges.

To keep an attribute $m \in \mathbb{F}_p$ hidden from the authorities, the users run \algorithm{PrepareBlindSign} to produce $\Lambda=(\gamma, c_m, c, \pi_{s})$. They create an \elgamal keypair $(d, \gamma=g_1^{d})$, pick a random  $o \in \mathbb{F}_p$, and compute a commitment $c_m=g_1^mh_1^o$. Then, the users compute $h=\hashtopoint(c_m)$ and  the encryption of $h^m$ as below:
\begin{equation}\nonumber
c = Enc(h^m) = (a, b) = (g_1^k,\gamma^kh^m),
\end{equation} 
where $k \in \mathbb{F}_p$. Finally, the users send $(\Lambda, \phi)$ to the signer, where $\pi_{s}$ is a zero-knowledge proof ensuring that $m$ satisfies the application-specific predicate $\phi$, and correctness of $\gamma, c_m,c$~(\ding{202}). All the zero-knowledge proofs required by \sysname are based on standard sigma protocols to show knowledge of representation of discrete logarithms; they are based on the DH assumption~\cite{camenisch1997proof} and do not require any trusted setup.

To blindly sign the attribute, each authority $i$ verifies the proof $\pi_{s}$, and uses the homomorphic properties of \elgamal to generate an encryption $\tilde{c}$ of $h^{x_i+y_i\cdot m}$ as below:
\begin{equation}\nonumber
\tilde{c} = (a^y, h^{x_i} b^{y_i}) = (g_1^{ky_i}, \gamma^{ky_i}h^{x_i+y_i\cdot m})
\end{equation}

Note that every authority must operate on the same element $h$. Intuitively, generating $h$ from $h=\hashtopoint(c_m)$ is equivalent to computing $h=g_1^{\tilde{r}}$  where $\tilde{r} \in \mathbb{F}_p$ is unknown by the users (as in Pointcheval and Sanders~\cite{pointcheval}). However, since $h$ is deterministic, every authority can uniquely derive it in isolation and forgeries are prevented since different $m_0$ and $m_1$ cannot lead to the same value of $h$.\footnote{If an adversary $\mathcal{A}$ can obtain two credentials $\sigma_0$ and $\sigma_1$ on respectively $m_0=0$ and $m_1=1$ with the same value $h$ as follows: $\sigma_0 = h^{x} \quad {\rm and} \quad \sigma_1=h^{x+y}$; then $\mathcal{A}$ could forge a new credential $\sigma_2$ on $m_2=2$:
$\sigma_2 = (\sigma_0)^{-1} \sigma_1 \sigma_1 = h^{x+2y}$.}
As described in \Cref{pointcheval_recall}, the blind signature scheme of Pointcheval and Sanders builds the credentials directly from a commitment of the attribute and a blinding factor secretly chosen by the authority; this is unsuitable for issuance of threshold credentials. We circumvent that problem by introducing the \elgamal ciphertext $c$ in our scheme and exploiting its homomorphism, as described above.

Upon reception of $\tilde{c}$, the users decrypt it using their \elgamal private key $d$ to recover the partial credentials $\sigma_i = (h, h^{x_i+y_i\cdot m})$; this is performed by the \algorithm{Unblind} algorithm (\ding{203}). Then, the users can call the \algorithm{AggCred} algorithm to aggregate any subset of $t$ partial credentials. This algorithm uses the Lagrange basis polynomial $l$ which allows to reconstruct the original $v(0)$ and $w(0)$ through polynomial interpolation;
\begin{equation}\nonumber
v(0) = \sum^t_{i=1} v(i)l_i \quad {\rm and} \quad w(0) = \sum^t_{i=1} w(i)l_i
\end{equation}
However, this computation happens in the exponent---neither the authorities nor the users should know the values $v(0)$ and $w(0)$. One can easily verify the correctness of \algorithm{AggCred} of $t$ partial credentials $\sigma_i=(h_i,s_i)$ as below.

\begin{eqnarray} \nonumber
	s &=& \prod^t_{i=1} \left(s_i\right)^{l_i} = \prod^t_{i=1} \left(h^{x_i+y_i\cdot m}\right)^{l_i} \\ \nonumber
	&=& \prod^t_{i=1} \left(h^{x_i}\right)^{l_i} \prod^t_{i=1} \left(h^{y_i\cdot m}\right)^{l_i} = \prod^t_{i=1} h^{(x_i l_i)} \prod^t_{i=1} h^{(y_i l_i)\cdot m} \\ \nonumber
	&=& h^{v(0)+w(0)\cdot m} = h^{x+y\cdot m}
\end{eqnarray}

Before verification, the verifier collects and aggregates the verifications keys of the authorities---this process  happens only once and ahead of time. The algorithms \algorithm{ProveCred} and \algorithm{VerifyCred} implement verification. First, the users randomize the credentials by picking a random $r' \in \mathbb{F}_p$ and computing $\sigma'=(h',s')=(h^{r'},s^{r'})$; then, they compute $\kappa$ and $\nu$ from the attribute $m$, a blinding factor $r\in\mathbb{F}_p$ and the aggregated verification key:
\begin{equation*}
\kappa=\alpha\beta^m g_2^r \qquad{\rm and}\qquad \nu=(h')^r
\end{equation*}
Finally, they send $\Theta=(\kappa, \nu, \sigma', \pi_v)$ and $\phi'$ to the verifier where $\pi_v$ is a zero-knowledge proof asserting the correctness of $\kappa$  and $\nu$; and that the private attribute $m$ embedded into $\sigma$ satisfies the application-specific predicate $\phi'$~(\ding{204}). The proof $\pi_v$ also ensures that the users actually know $m$ and that $\kappa$ has been built using the correct verification keys and blinding factors. The pairing verification is similar to Pointcheval and Sanders~\cite{pointcheval} and Boneh~\etal~\cite{bls}; expressing $h'=g_1^{\tilde{r}} \; | \; \tilde{r} \in \mathbb{F}_p$, the left-hand side of the pairing verification can be expanded as:
\begin{equation*}
    e(h',\kappa) = e(h',g_2^{(x+my+r)}) = e(g_1,g_2)^{(x+my+r) \tilde{r}}
\end{equation*}
and the right-hand side:
\begin{equation*}
    e(s' \nu,g_2) = e(h'^{(x+my+r)},g_2) = e(g_1,g_2)^{(x+my+r)  \tilde{r}}
\end{equation*}
From where the correctness of \algorithm{VerifyCred} follows.

\paragraph{Security}

The proof system we require is based on standard sigma protocols to show knowledge of representation of discrete logarithms, and can be rendered non-interactive using the Fiat-Shamir heuristic~\cite{fiat1986prove} in the random oracle model.  As our signature scheme is derived from the ones due to Pointcheval and Sanders~\cite{pointcheval} and BLS~\cite{bls}, we inherit their assumptions as well; namely, LRSW~\cite{lysyanskaya1999pseudonym} and XDH~\cite{bls}.

\begin{theorem} \label{th:coconut_theorem}
Assuming LRSW, XDH, and the existence of random oracles, \sysname is a secure threshold credentials scheme, meaning it satisfies unforgeability (as long as fewer than $t$ authorities collude), blindness, and unlinkability.
\end{theorem}
\noindent A sketch of this proof, based on the security of the underlying components of \sysname, can be found in Appendix~\ref{security_proofs}. \sysname guarantees unforgeability as long as less than $t$ authorities collude ($t>n/2$), and guarantees blindness and unlinkability no matter how many authorities collude (and even if the verifier colludes with the authorities).

\subsection{Multi-Attribute Credentials}\label{multi_message_scheme}
We expand our scheme to embed multiple attributes into a single credential without increasing its size; this generalization follows directly from the Waters signature scheme~\cite{waters} and Pointcheval and Sanders~\cite{pointcheval}. The authorities' key pairs become:
\begin{equation}\nonumber
sk= (x,y_1,\dots,y_{q}) \quad{\rm and}\quad vk = (g_2,g_2^x,g_2^{y_1}, \dots, g_2^{y_{q}}) 
\end{equation}
where $q$ is the number of attributes. The multi-attribute credential is derived from the commitment $c_m$ and the group element $h$ as below:
\begin{equation*}
c_m = g_1^o \prod_{j=1}^{q} h_j^{m_j} \qquad{\rm and}\qquad h = \hashtopoint(c_m)
\end{equation*}
and the credential generalizes as follows:
\begin{equation*}
\sigma = (h,h^{x+\sum_{j=1}^{q} m_j y_j})
\end{equation*}
The credential's size does not increase with the number of attributes or authorities---it is always composed of two group elements. The security proof of the multi-attribute scheme relies on a reduction against the single-attribute scheme and is analogous to Pointcheval and Sanders~\cite{pointcheval}. Moreover, it is also possible to combine public and private attributes to keep only a subset of the attributes hidden from the authorities, while revealing some others; the \algorithm{BlindSign} algorithm only verifies the proof $\pi_{s}$ on the private attributes (similar to Chase~\etal~\cite{amac}). The full primitives of the multi-attribute cryptographic scheme are presented in Appendix~\ref{multi_attribute_credentials}.

If the credentials include only non-random attributes, the verifier could guess its value by brute-forcing the verification algorithm\footnote{Let assume for example that some credentials include a single attribute $m$ representing the age of the user; the verifier can run the verification algorithm $e(h,\kappa(\alpha\cdot\beta^m)^{-1})=e(\nu,g_2)$ for every $m \in [1,100]$ and guess the value of $m$.}. This issue is prevented by always embedding a private random attribute into the credentials, that can also act as the authorization key for the credential.


%% file: sections/implementation.tex
\section{Implementation} 
\label{implementation}

We implement a Python library for \sysname as described in \Cref{construction} and publish the code on GitHub as an open-source project\footnote{\url{https://github.com/asonnino/coconut}}.
We also implement a smart contract library in Chainspace~\cite{chainspace} to enable other application-specific smart contracts (see \Cref{applications}) to conveniently use our cryptographic primitives.
We present the design and implementation of the \sysname smart contract library in Section~\ref{smart_contract_library}.
In addition, we implement and evaluate some of the functionality of the \sysname smart contract library in \ethereum~\cite{ethereum} (Section~\ref{ethereum_smart_contract_library}).
 Finally, we show how to integrate \sysname into existing semi-permissioned \blockchains (Section~\ref{integrations_into_ledgers}).

\subsection{The \sysname Smart Contract Library} 
\label{smart_contract_library}

\begin{figure}[t]
    \centering
    \includegraphics[width=.48\textwidth]{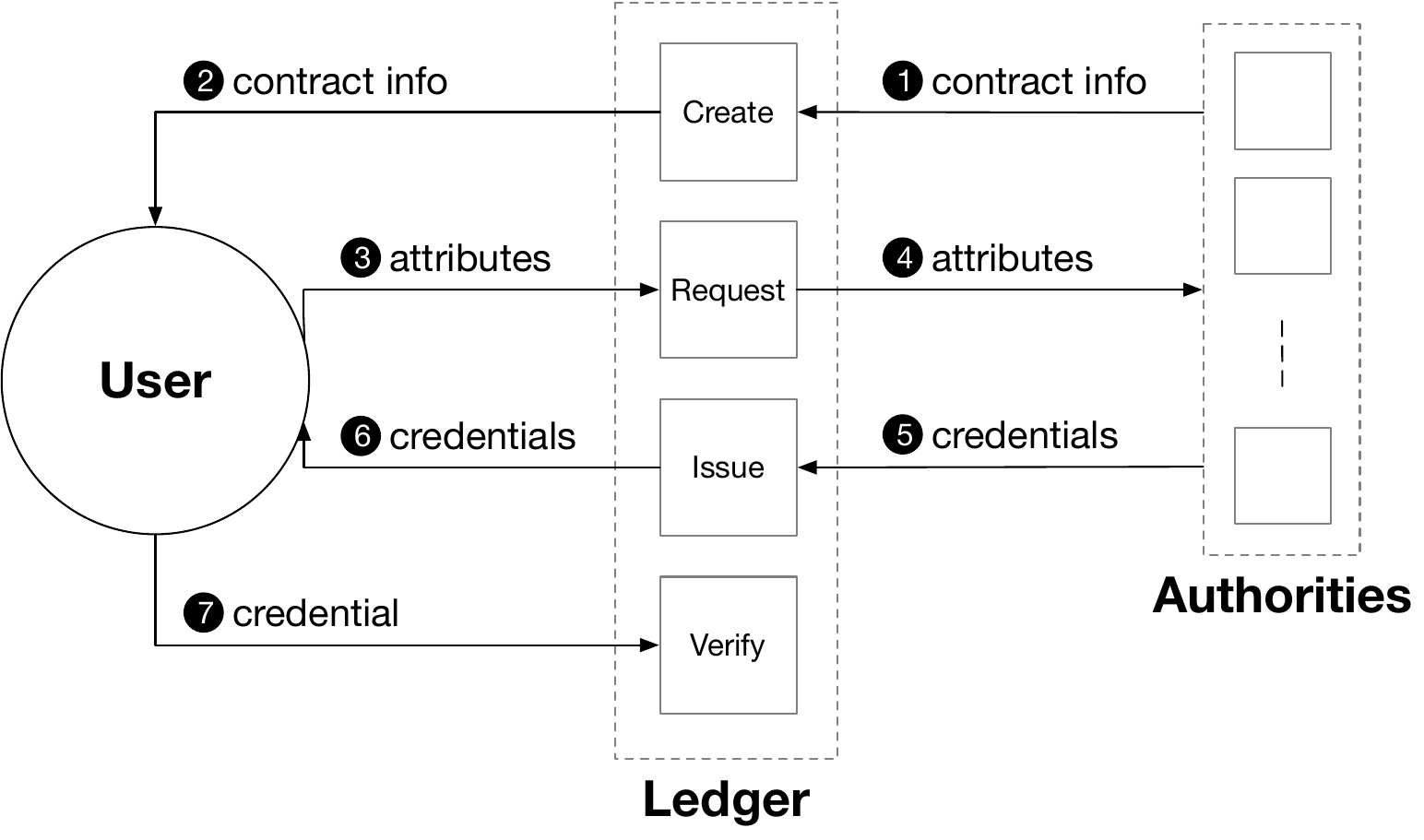}
    \caption{\footnotesize The \sysname smart contract library.}
    \label{fig:library}
\end{figure}

We implement the \sysname smart contract in \chainspace\footnote{\url{https://github.com/asonnino/coconut-chainspace}} (which can be used by other application-specific smart contracts) as a library to issue and verify randomizable threshold credentials through cross-contract calls. 
The contract has four functions, \algorithm{(Create, Request, Issue, Verify)}, as illustrated in \Cref{fig:library}. First, a set of authorities call the \algorithm{Create} function to initialize a \sysname instance defining the \emph{contract info}; i.e., their verification key, the number of authorities and the threshold parameter~(\ding{202}). The initiator smart contract can specify a callback contract that needs to be executed by the user in order to request credentials; e.g., this callback can be used for authentication. The instance is public and can be read by the user~(\ding{203}); any user can request a credential through the \algorithm{Request} function by executing the specified callback contract, and providing the public and private \emph{attributes} to include in the credentials~(\ding{204}). The public attributes are simply a list of clear text strings, while the private attributes are encrypted as described in \Cref{threshold_credentials_scheme}. Each signing authority monitors the \blockchain at all times, looking for credential requests. If the request appears on the \blockchain (i.e., a transaction is executed), it means that the callback has been correctly executed~(\ding{205}); each authority issues a partial \emph{credential} on the specified attributes by calling the \algorithm{Issue} procedure~(\ding{206}). In our implementation, all partial credentials are in the \blockchain; however, these can also be provided to the user off-chain. Users collect a threshold number of partial credentials, and aggregate them to form a full credential~(\ding{207}). Then, the users locally randomize the credential. The last function of the \sysname library contract is \algorithm{Verify} that allows the \blockchain---and anyone else---to check the validity of a given credential~(\ding{208}). 

A limitation of this architecture is that it is not efficient for the authorities to continuously monitor the \blockchain. \Cref{integrations_into_ledgers} explains how to overcome this limitation by embedding the authorities into the nodes running the \blockchain.

\subsection{Ethereum Smart Contract Library}\label{ethereum_smart_contract_library}

To make \sysname more widely available, we also implement it in \ethereum---a popular permissionless smart contract \blockchain~\cite{ethereum}.
We release the \sysname \ethereum smart contract as an open source library\footnote{\url{https://github.com/musalbas/coconut-ethereum}}.
The library is written in Solidity, a high-level JavaScript-like language that compiles down to \ethereum Virtual Machine (EVM) assembly code.
\ethereum recently hardcoded a pre-compiled smart contract in the EVM for performing pairing checks and elliptic curve operations on the alt\_bn128 curve~\cite{eip197,eip196}, for efficient verification of zkSNARKs. The execution of an \ethereum smart contract has an associated `gas cost', a fee that is paid to miners for executing a transaction. Gas cost is calculated based on the operations executed by the contract; i.e.,\ the more operations, the higher the gas cost. Pre-compiled contracts have lower gas costs than equivalent native smart contracts.

We use the pre-compiled contract for performing a pairing check, in order to implement \sysname verification within a smart contract. The \ethereum code only implements elliptic curve addition and scalar multiplication on $\mathbb{G}_1$, whereas \sysname requires operations on $\mathbb{G}_2$ to verify credentials. 
Therefore, we implement elliptic curve addition and scalar multiplication on $\mathbb{G}_2$ as an \ethereum smart contract library written in Solidity that we also release open source\footnote{\url{https://github.com/musalbas/solidity-BN256G2}}. This is a practical solution for many \sysname applications, as verifying credentials with one revealed attribute only requires one addition and one scalar multiplication. It would not be practical however to verify credentials with attributes that will not be revealed---this requires three $\mathbb{G}_2$ multiplications using our elliptic curve implementation, which would exceed the current \ethereum block gas limit (8M as of February 2018).

We can however use the \ethereum contract to design a federated peg for side chains, or a coin tumbler as an \ethereum smart contract, based on credentials that reveal one attribute. We go on to describe and implement this tumbler using the \sysname \chainspace library in \Cref{tumbler}, however the design for the \ethereum version differs slightly to avoid the use of attributes that will not be revealed, which we describe in Appendix~\ref{eth-tumbler}. The library shares the same functions as the \chainspace library described in \Cref{smart_contract_library}, except for \algorithm{Request} and \algorithm{Issue} which are computed off the blockchain to save gas costs.
As \algorithm{Request} and \algorithm{Issue} functions simply act as a communication channel between users and authorities, users can directly communicate with authorities off the \blockchain to request tokens.
 This saves significant gas costs that would be incurred by storing these functions on the \blockchain. The \algorithm{Verify} function simply verifies tokens against \sysname instances created by the \algorithm{Create} function. 

\subsection{Deeper Blockchain Integration} \label{integrations_into_ledgers}

The designs described in \Cref{smart_contract_library} and \Cref{ethereum_smart_contract_library} rely on authorities on-the-side for issuing credentials. In this section, we present designs that incorporate \sysname authorities within the infrastructure of a number of semi-permissioned \blockchains. This enables the issuance of credentials as a side effect of the normal system operations, taking no additional dependency on extra authorities. It remains an open problem how to embed \sysname into permissionless systems, based on proof of work or stake. These systems have a highly dynamic set of nodes maintaining the state of their blockchains, which cannot readily be mapped into \sysname issuing authorities.

Integration of \sysname into \hyperledger Fabric~\cite{hyperledger}---a permissioned \blockchain platform---is straightforward. Fabric contracts run on private sets of computation nodes---and use the Fabric protocols for cross-contract calls. In this setting, \sysname issuing authorities can coincide with the Fabric smart contract authorities. Upon a contract setup, they perform the setup and key distribution, and then issue partial credentials when authorized by the contract. For issuing \sysname credentials, the only secrets maintained are the private issuing keys; all other operations of the contract can be logged and publicly verified. \sysname has obvious advantages over using traditional CL credentials relying on a single authority---as currently present in the \hyperledger roadmap\footnote{\url{http://nick-fabric.readthedocs.io/en/latest/idemix.html}}. The threshold trust assumption---namely that integrity and availability is guaranteed under the corruption of a subset of authorities is preserved, and prevents forgeries by a single corrupted node.

We can also naturally embed \sysname into sharded scalable \blockchains, as exemplified by \chainspace~\cite{chainspace} (which supports general smart contracts), and \omniledger~\cite{kokoris2017omniledger} (which supports digital tokens). 
In these systems, transactions are distributed and executed on `shards' of authorities, whose membership and public keys are known. 
\sysname authorities can naturally coincide with the nodes within a shard---a special transaction type in \omniledger, or a special object in \chainspace, can signal to them that issuing a credential is authorized. 
The authorities,  then issue the partial signature necessary to reconstruct the \sysname credential, and attach it to the transaction they are processing anyway. 
Users can aggregate, re-randomize and show the credential. 


%% file: sections/applications.tex
\section{Applications} \label{applications}

In this section, we present three applications that leverage \sysname to offer improved security and privacy properties---a coin tumbler (Section~\ref{tumbler}), a privacy-preserving petition system (Section~\ref{petition}), and a system for censorship-resistant distribution of proxies (Section~\ref{proxy}).  
For generality, the applications assume authorities external to the \blockchain, but these can also be embedded into the \blockchain as described in \Cref{integrations_into_ledgers}.

\subsection{Coin Tumbler} 
\label{tumbler}

\begin{figure}[t]
    \centering
    \includegraphics[width=.48\textwidth]{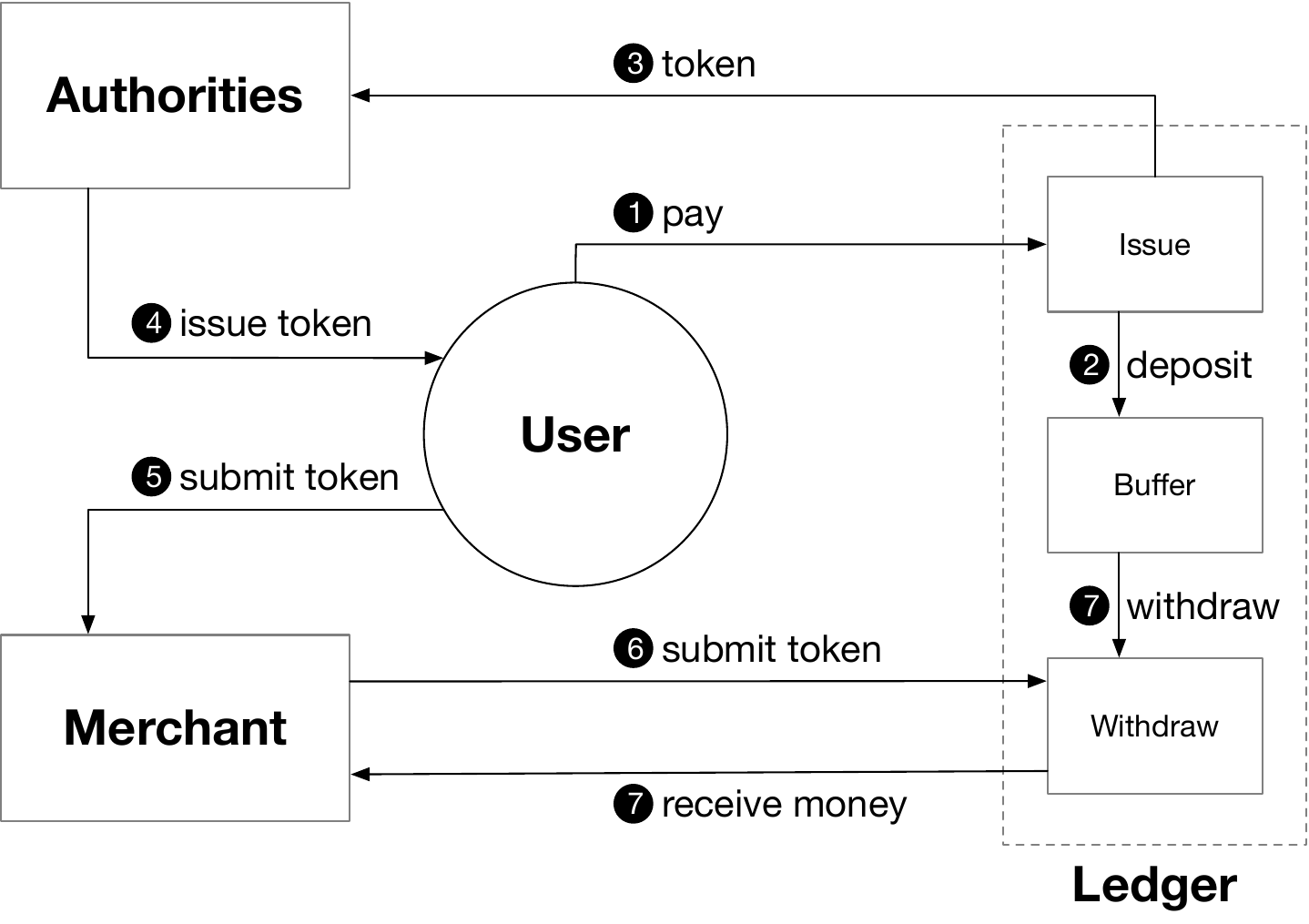}
    \caption{\footnotesize The coin tumbler application.}
    \label{fig:tumbler}
\end{figure}

We implement a coin tumbler (or mixer) on \chainspace as depicted in \Cref{fig:tumbler}. Coin tumbling is a method to mix cryptocurrency associated with an address visible in a public ledger with other addresses, to ``clean'' the coins and obscure the trail back to the coins' original source address. A limitation of previous similar schemes~\cite{mixcoin,blindcoin,tumblebit,coinjoin,coinshuffle,xim,mobius}  
 is that they are either centralized (i.e., there is a central authority that operates the tumbler, which may go offline), or require users to coordinate with each other.
The \sysname tumbler addresses these issues \via a distributed design (i.e., security relies on a set of multiple authorities that are collectively trusted to contain at least $t$ honest ones), and does not require users to coordinate with each other.
Zcash~\cite{zcash} achieves a similar goal: it theoretically hides the totality of the transaction but at a large computational cost, and offers the option to cheaply send transactions in clear. In practice, the computational overhead of sending hidden transactions makes it impractical, and only a few users take advantage of the optional privacy provided by Zcash; as a result, transactions are easy to de-anonymize~\cite{Kappos}, and recent works aim to reduce the computational overhead of Zcash hidden transactions~\cite{fast-zcash}. \sysname provides efficient proofs taking only a few milliseconds (see \Cref{evaluation}), and makes hidden transactions practical. Trust assumptions in Zcash are different from \sysname. However, instead of assuming a threshold number of honest authorities, Zcash relies on zk-SNARKs which assumes a setup algorithm executed by a trusted authority\footnote{Recent proposals aim to distribute this trusted setup~\cite{mpc-zcash-1}.}.
M{\"o}bius~\cite{mobius}---which was developed concurrently---is a coin tumbler based on \ethereum smart contracts that achieves strong notions of anonymity and low off-chain communication complexity. M{\"o}bius relies on ring signatures to allow parties to prove group membership without
revealing exactly which public key belongs to them.

Our tumbler uses \sysname to instantiate a pegged side-chain~\cite{back2014enabling}, providing stronger value transfer anonymity than the original cryptocurrency platform, through unlinkability between issuing a credential representing an e-coin~\cite{chaum1988untraceable}, and spending it. The tumbler application is based on the \sysname contract library and an application specific smart contract called ``tumbler''. 

A set of authorities jointly create an instance of the \sysname smart contract as described in \Cref{smart_contract_library} and specify the smart contract handling the coins of the underlying \blockchain as callback. Specifically, the callback requires a coin transfer to a buffer account. Then users execute the callback and \emph{pay} $v$ coins to the buffer to ask a credential on the public attribute $v$, and on two private attributes: the user's private key $k$ and a randomly generated sequence number $s$~(\ding{202}). Note that to prevent tracing traffic analysis, $v$ should be limited to a specific set of possible values (similar to cash denominations). The request is accepted by the \blockchain only if the user \emph{deposited} $v$ coins to the buffer account~(\ding{203}). 

Each authority monitors the \blockchain and detects the \emph{request}~(\ding{204}); and issues a partial \emph{credential} to the user (either on chain or off-chain)~(\ding{205}). The user aggregates all partial credentials into a consolidated credential, re-randomizes it, and \emph{submits} it as coin token to a merchant. First, the user produces a zk-proof of knowledge of its private key by binding the proof to the merchant's address $addr$; then, the user provides the merchant with the proof along with the sequence number $s$ and the consolidated credential~(\ding{206}). The coins can only be spent with knowledge of the associated sequence number and by the owner of $addr$. 
To accept the above as payment, the merchant \emph{submits} the token by showing the credential and a group element $\zeta=g_1^s\in\mathbb{G}_1$ to the tumbler contract along with a zero-knowledge proof ensuring that $\zeta$ is well-formed~(\ding{207}). To prevent double spending, the tumbler contract keeps a record of all elements $\zeta$ that have already been shown. Upon showing a $\zeta$ embedding a fresh (unspent) sequence number $s$, the contract verifies that the credential and zero-knowledge proofs check, and that $\zeta$ doesn't already appear in the spent list. Then it \emph{withdraws} $v$ coins from the buffer~(\ding{208}), sends them to be \emph{received} by the merchant account determined by $addr$, and adds $\zeta$ to the spent list~(\ding{209}). For the sake of simplicity, we keep the transfer value $v$ in clear-text (treated as a public attribute), but this could be easily hidden by integrating a range proof; this can be efficiently implemented using the technique developed by B{\"u}nz~\etal~\cite{bunzbulletproofs}.

\noindent\textbf{Security consideration.} \sysname provides blind issuance which allows the user to obtain a credential on the sequence number $s$ without the authorities learning its value. Without blindness, any authority seeing the user key $k$ could potentially race the user and the merchant, and spend it---blindness prevents authorities from stealing the token. Furthermore, \sysname provides unlinkability between the \emph{pay} phase~(\ding{202}) and the \emph{submit} phase~(\ding{206}) (see \Cref{fig:tumbler}), and prevents any authority or third parties from keeping track of the user's transactions. As a result, a merchant can receive payments for good or services offered, yet not identify the purchasers. Keeping a spent list of all elements $\zeta$ prevents double-spending attacks~\cite{karame2012double} without revealing the sequence number $s$; this prevents an attacker from exploiting a race condition in the \emph{submit token} phase~(\ding{207}) and lock user's funds\footnote{An attacker observing a sequence number $s$ during a \emph{submit token} phase~(\ding{207}) could exploit a race condition to lock users fund by quickly buying a token using the same $s$, and spending it before the original \emph{submit token} phase is over.}.
Finally, this application prevents a single authority from creating coins to steal all the money in the buffer. The threshold property of \sysname implies that the adversary needs to corrupt at least $t$ authorities for this attack to be possible. A small subset of authorities cannot block the issuance of a token---the service is guaranteed to be available as long as at least $t$ authorities are running. 

\subsection{Privacy-preserving petition} \label{petition}

\begin{figure}[t]
    \centering
    \includegraphics[width=.48\textwidth]{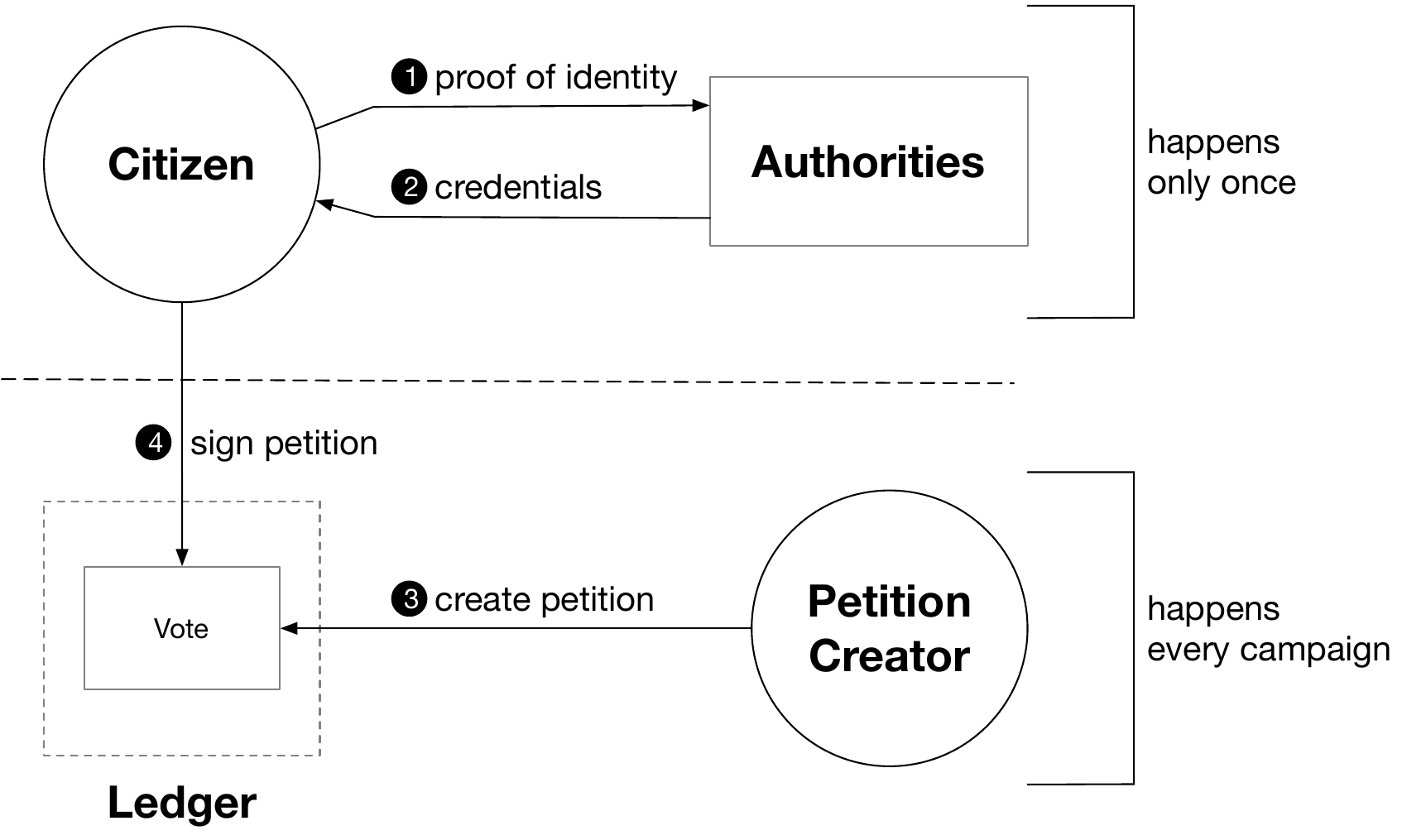}
    \caption{\footnotesize The petition application.}
    \label{fig:petition}
\end{figure}

We consider the scenario where several authorities managing the country \emph{C} wish to issue some long-term credentials to its citizens to enable any third party to organize a privacy-preserving petition.  All citizens of \emph{C} are allowed to participate, but should remain anonymous and unlinkable across petitions. This application extends the work of Diaz~\etal~\cite{diaz2008privacy} which does not consider threshold issuance of credentials.

Our petition system is based on the \sysname library contract and a simple smart contract called ``petition".
There are three types of parties: a set of signing authorities representing \emph{C}, a petition initiator, and the citizens of \emph{C}.  
The signing authorities create an instance of the \sysname smart contract as described in \Cref{smart_contract_library}. As shown in \Cref{fig:petition}, the citizen provides a \emph{proof of identity} to the authorities~(\ding{202}). 
 The authorities check the citizen's identity, and issue a blind and long-term signature on her private key $k$.
 This signature, which the citizen needs to obtain only once, acts as her long term \emph{credential} to sign any petition~(\ding{203}).
 
Any third party can \emph{create a petition} by creating a new instance of the petition contract and become the ``owner" of the petition. The petition instance specifies an identifier $g_s \in \mathbb{G}_1$ unique to the petition where its representation is unlinkable to the other points of the scheme\footnote{This identifier can be generated through a hash function $\mathbb{F}_p \rightarrow \mathbb{G}_1:\widetilde{H}(s)=g_s \;|\; s\in\mathbb{F}_p$.}, as well as the verification key of the authorities issuing the credentials and any application specific parameters (e.g., the options and current votes)~(\ding{204}). In order to \emph{sign} a petition, the citizens compute a value $\zeta=g_s^{k}$. They then adapt the zero-knowledge proof of the \algorithm{ProveCred} algorithm of \Cref{threshold_credentials_scheme} to show that $\zeta$ is built from the same attribute $k$ in the credential; the petition contract checks the proofs  and the credentials, and checks that the signature is fresh by verifying that $\zeta$ is not part of a spent list. If all the checks pass, it adds the citizens' signatures to a list of records and adds $\zeta$ to the spent list to prevents a citizen from signing the same petition multiple times (prevent double spending)~(\ding{205}). Also, the zero-knowledge proof  ensures that $\zeta$ has been built from a signed private key $k$; this means that the users correctly executed the callback to prove that they are citizens of \emph{C}. 

\noindent\textbf{Security consideration.} \sysname's blindness property prevents the authorities from learning the citizen's secret key, and misusing it to sign petitions on behalf of the citizen. 
Another benefit is that it lets citizens sign petitions anonymously; citizens only have to go through the issuance phase once, and can then re-use credentials multiple times while staying anonymous and unlinkable across petitions. \sysname allows for distributed credentials issuance, removing a central authority and preventing a single entity from creating arbitrary credentials to sign petitions multiple times. 

\subsection{Censorship-resistant distribution of proxies} 
\label{proxy}

\begin{figure}[t]
    \centering
    \includegraphics[width=.48\textwidth]{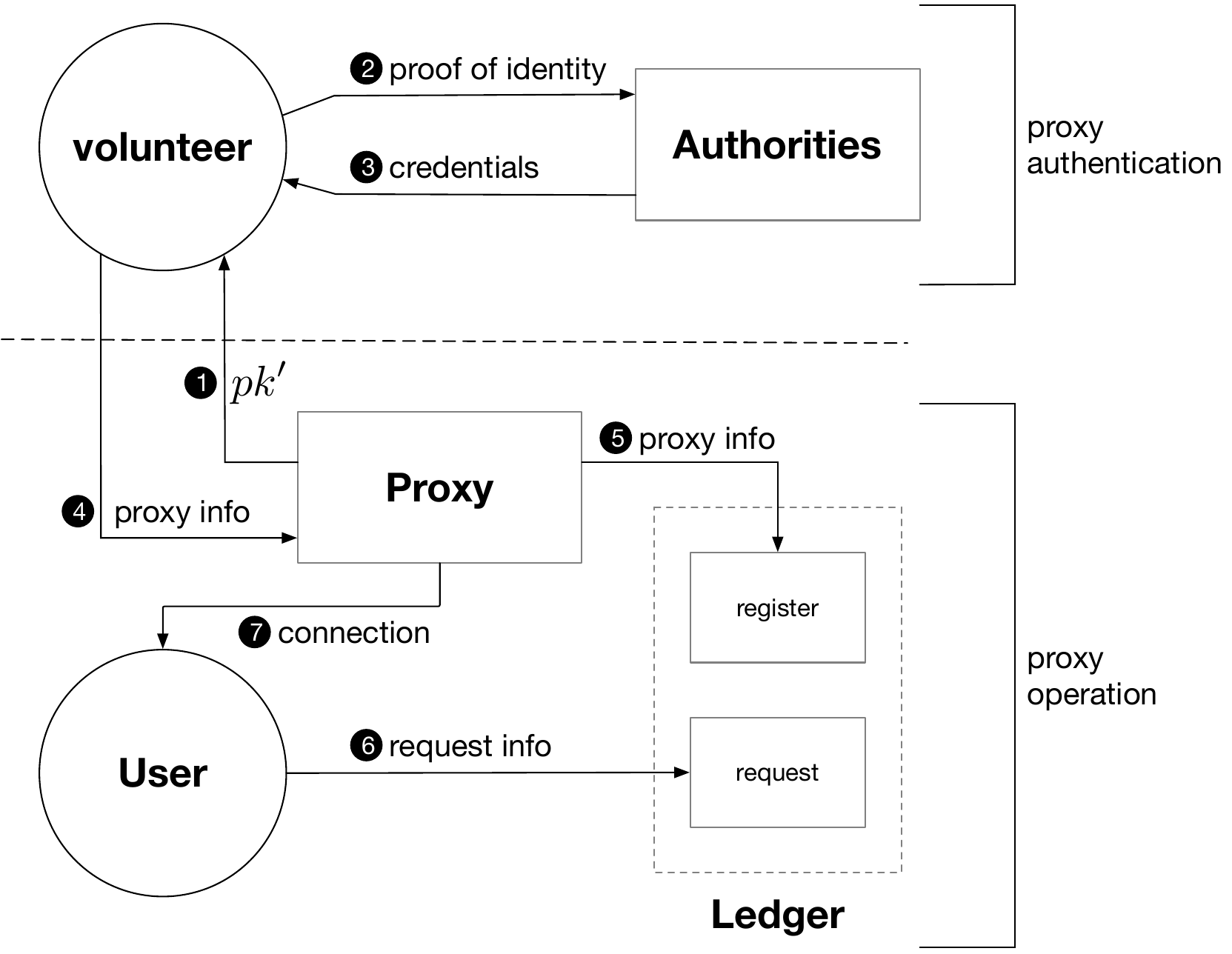}
    \caption{\footnotesize The censorship-resistant proxy distribution system.}
    \label{fig:proxy}
\end{figure}

Proxies can be used to bypass censorship, but often become the target of censorship themselves. We present a system based on \sysname for censorship-resistant distribution of proxies (CRS). In our CRS, the volunteer \emph{V} runs proxies, and is known to the \sysname authorities through its long-term public key. The authorities establish reputability of volunteers (identified by their public keys) through an out of band mechanism. The user \emph{U} wants to find proxy IP addresses belonging to reputable volunteers, but volunteers want to hide their identity. As shown in~\Cref{fig:proxy}, \emph{V} gets an ephemeral public key $pk'$ from the proxy~(\ding{202}), provides \emph{proof of identity} to the authorities~(\ding{203}), and gets a \emph{credential} on two private attributes: the proxy IP address, $pk'$, and the time period $\delta$ for which it is valid~(\ding{204}).

\emph{V} shares the credential with the concerned proxy~(\ding{205}), which creates the \emph{proxy info} including $pk'$, $\delta$, and the credential; the proxy `registers' itself by appending this information to the \blockchain along with a zero-knowledge proof and the material necessary to verify the validity of the credential~(\ding{206}).

The users \emph{U} monitor the \blockchain for proxy registrations. When a registration is found, \emph{U} indicates the intent to use a proxy by publishing to the \blockchain a \emph{request info} message which looks as follows: user IP address encrypted under $pk'$ which is embedded in the registration \blockchain entry~(\ding{207}).  
The proxy continuously monitors the \blockchain, and upon finding a user request
addressed to itself, \emph{connects} to \emph{U} and presents proof of knowledge of the private key associated with $pk'$~(\ding{208}). 
\emph{U} verifies the proof, the proxy IP address and its validity period, and then starts relaying its traffic through the proxy.

\noindent\textbf{Security consideration.}
A common limitation of censorship resistance schemes is relying on volunteers that are \emph{assumed} to be resistant to coercion: either \first the volunteer is a large, commercial organisation (e.g., Amazon or Google) over which the censor cannot exert its influence; and/or \second the volunteer is located outside the country of censorship. 
However, both these assumptions were proven wrong~\cite{meek-suspended,five-eyes}. 
The proposed CRS overcomes this limitation by offering coercion-resistance to volunteers from censor-controlled users and authorities. Due to \sysname's blindness property, a volunteer can get a credential on its IP address and ephemeral public key without revealing those to the authorities. The users get proxy IP addresses run by the volunteer, while being unable to link it to the volunteer's long-term public key.
Moreover, the authorities operate independently and can be controlled by different entities, and are resilient against a threshold number of authorities being dishonest or taken down. 

%% file: sections/evaluation.tex
\section{Evaluation} 
\label{evaluation}
We present the evaluation of the \sysname threshold credentials scheme; first we present a benchmark of the cryptographic primitives described in \Cref{construction} and then we evaluate the smart contracts described in \Cref{applications}.

\subsection{Cryptographic Primitives}
We implement the primitives described in Section~\ref{construction} in Python using petlib\footnote{\url{https://github.com/gdanezis/petlib}} and bplib\footnote{\url{https://github.com/gdanezis/bplib}}. The bilinear pairing is defined over the Barreto-Naehrig~\cite{Barreto-Naehrig} curve, using OpenSSL as arithmetic backend.


\begin{table}[t]
\begin{center}
    \begin{tabular}{lcc}
        \toprule
        \small\bf Operation & \small\bf \boldmath $\mu$ [ms] & \small\bf \boldmath$\sqrt{\sigma^2}$ [ms]\\
        \midrule
        \textsf{PrepareBlindSign} & 2.633 & $\pm$ 0.003 \\ 
        \textsf{BlindSign} & 3.356 & $\pm$ 0.002 \\
         \textsf{Unblind} & 0.445 & $\pm$ 0.002 \\ 
          \textsf{AggCred} & 0.454 & $\pm$ 0.000 \\
        \textsf{ProveCred} & 1.544 & $\pm$ 0.001 \\
        \textsf{VerifyCred} & 10.497 & $\pm$ 0.002 \\
        \bottomrule
    \end{tabular}
\end{center}
\caption{\footnotesize Execution times for the cryptographic primitives described in \Cref{construction}, measured for one private attribute over 10,000 runs. \textsf{AggCred} is computed assuming two authorities; the other primitives are independent of the number of authorities.}
\label{tab:timing}
\end{table}

\begin{table}[t]
\begin{center}
    \begin{tabular}{l c c}
        \toprule
        \multicolumn{3}{l}{Number of authorities: $n$, Signature size: 132 bytes}\\
        \small\bf Transaction & \small\bf complexity & \small\bf size [B]\\
        \midrule
        \multicolumn{1}{l}{Signature on one public attribute:}\\
        \ding{202} request credential & $O(n)$ & 32 \\ 
        \ding{203} issue credential &  $O(n)$ & 132 \\
        \ding{204} verify credential &  $O(1)$ &  162 \\ 
        &&\\
        \multicolumn{1}{l}{Signature on one private attribute:}\\
        \ding{202} request credential &  $O(n)$ & 516 \\ 
        \ding{203} issue credential &  $O(n)$ & 132 \\
        \ding{204} verify credential &  $O(1)$ &  355 \\
       \bottomrule
    \end{tabular}
\end{center}
\caption{\footnotesize Communication complexity and transaction size for the \sysname credentials scheme when signing one public and one private attribute (see \Cref{fig:protocol_priv} of \Cref{construction}).}
\label{tab:transactions}
\end{table}

\paragraph{Timing benchmark} \Cref{tab:timing} shows the mean ($\mu$) and standard deviation ($\sqrt{\sigma^2}$) of the execution of each procedure described in section \Cref{construction}. Each entry is the result of 10,000 runs measured on an Octa-core Dell desktop computer, 3.6GHz Intel Xeon. Signing is much faster than verifying credentials---due to the pairing operation in the latter; verification takes about 10ms; signing a private attribute is about 3 times faster.


\paragraph{Communication complexity and packets size} \Cref{tab:transactions} shows the communication complexity and the size of each exchange involved in the \sysname credentials scheme, as presented in \Cref{fig:protocol_priv}. The communication complexity is expressed as a function of the number of signing authorities ($n$), and the size of each attribute is limited to 32 bytes as the output of the SHA-2 hash function. The size of a credential is 132 bytes. The highest transaction sizes are to request and verify credentials embedding a private attribute; this is due to the proofs $\pi_s$ and $\pi_v$ (see \Cref{construction}). The proof $\pi_s$ is approximately 318 bytes and $\pi_v$ is 157 bytes.

\begin{figure}[t]
    \centering
    \includegraphics[width=.45\textwidth]{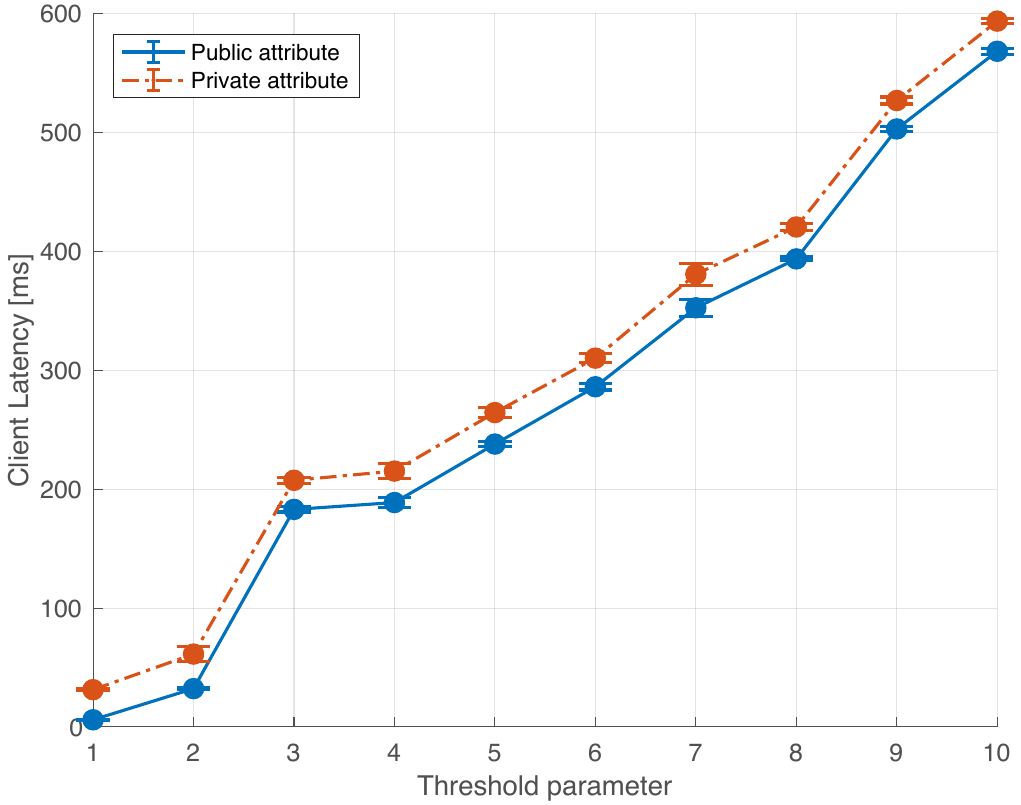}
    \caption{\footnotesize Client-perceived latency for \sysname threshold credentials scheme with geographically distributed authorities, measured for one attribute over 100 runs. }
    \label{fig:latency}
\end{figure}

\paragraph{Client-perceived latency} We evaluate the client-perceived latency for the \sysname threshold credentials scheme for authorities deployed on Amazon AWS~\cite{aws} when issuing partial credentials on one public and one private attribute. The client requests a partial credential from 10 authorities, and latency is defined as the time it waits to receive $t$-out-of-10 partial signatures. \Cref{fig:latency} presents measured latency for a threshold parameter t ranging from 1--10. 
The dots correspond to the average latency and the error-bars represent the normalized standard deviation, computed over 100 runs. The client is located in London while the 10 authorities are geographically distributed across the world; US East (Ohio), US West (N. California), Asia Pacific (Mumbai), Asia Pacific (Singapore), Asia Pacific (Sydney), Asia Pacific (Tokyo), Canada (Central), EU (Frankf{\"u}rt), EU (London), and South America (S{\~a}o Paulo). All machines are running a fresh 64-bit Ubuntu distribution, the client runs on a \emph{large} AWS instance and the authorities run on \emph{nano} instances.

As expected, we observe that the further the authorities are from the client, the higher the latency due to higher response times; the first authorities to respond are always those situated in Europe, while Sidney and Tokyo are the latest. 
Latency grows linearly, with the exception of a large jump (of about 150 ms) when $t$ increases from 2 to 3---this is due to the 7 remaining authorities being located outside Europe. 
The latency overhead between credential requests on public and private attributes remains constant. 

\begin{table}[t]
\begin{center}
    \begin{tabular}{lccc}
        \toprule
        \multicolumn{4}{l}{\sysname smart contract library}\\
        \small\bf Operation & \small\bf \boldmath $\mu$ [ms] & \small\bf \boldmath$\sqrt{\sigma^2}$ [ms] & \bf size [kB]\\
        \midrule
        \textsf{Create} [g] & 0.195 & $\pm$ 0.065 & $\sim1.38$\\ 
        \textsf{Create} [c] & 12.099 & $\pm$ 0.471 & -\\ 
        \textsf{Request} [g] & 7.094 & $\pm$ 0.641 & $\sim3.77$\\ 
        \textsf{Request} [c] & 6.605 & $\pm$ 0.559 & -\\ 
        \textsf{Issue} [g] & 4.382 & $\pm$ 0.654 & $\sim3.08$\\ 
        \textsf{Issue} [c] & 0.024 & $\pm$ 0.001 & -\\  
        \textsf{Verify} [g] & 5.545 & $\pm$ 0.859 & $\sim1.76$\\ 
        \textsf{Verify} [c] & 10.814 & $\pm$ 1.160 & - \\
        \bottomrule
    \end{tabular}
\end{center}
\caption{\footnotesize Timing and transaction size of the \chainspace implementation of the \sysname smart contract library described in \Cref{smart_contract_library}, measured for two authorities and one private attributes over 10,000 runs. The notation [g] denotes the execution the procedure and [c] denotes the execution of the checker.}
\label{tab:library}
\end{table}

\begin{table}[t]
\begin{center}
    \begin{tabular}{lccc}
        \toprule
        \multicolumn{4}{l}{Coin tumbler}\\
        \small\bf Operation & \small\bf \boldmath $\mu$ [ms] & \small\bf \boldmath$\sqrt{\sigma^2}$ [ms] & \bf size [kB]\\
        \midrule
        \textsf{InitTumbler} [g] & 0.235 & $\pm$ 0.065 & $\sim1.38$\\ 
        \textsf{InitTumbler} [c] & 19.359 & $\pm$ 0.773 & -\\ 
        \textsf{Pay} [g] & 11.939 & $\pm$ 0.792 & $\sim4.28$\\ 
        \textsf{Pay} [c] & 6.625 & $\pm$ 0.559 & -\\ 
        \textsf{Redeem} [g] & 0.132 & $\pm$ 0.012 & $\sim3.08$\\ 
        \textsf{Redeem} [c] & 11.742 & $\pm$ 0.757 & -\\  
        \bottomrule
    \end{tabular}
\end{center}
\caption{\footnotesize Timing and transaction size of the \chainspace implementation of the coin tumbler smart contract (described in \Cref{tumbler}), measured over 10,000 runs. The transactions are independent of the number of authorities.  The notation [g] denotes the execution the procedure and [c] denotes the execution of the checker.}
\label{tab:tumbler}
\end{table}

\begin{table}[t]
\begin{center}
    \begin{tabular}{lccc}
        \toprule
        \multicolumn{4}{l}{Privacy-preserving e-petition}\\
        \small\bf Operation & \small\bf \boldmath $\mu$ [ms] & \small\bf \boldmath$\sqrt{\sigma^2}$ [ms] & \bf size [kB]\\
        \midrule
        \textsf{InitPetition} [g] & 3.260 & $\pm$ 0.209 & $\sim1.50$\\ 
        \textsf{InitPetition} [c] & 3.677 & $\pm$ 0.126 & -\\ 
        \textsf{SignPetition} [g] & 7.999 & $\pm$ 0.467 & $\sim3.16$\\ 
        \textsf{SignPetition} [c] & 15.801 & $\pm$ 0.537 & -\\ 
        \bottomrule
    \end{tabular}
\end{center}
\caption{\footnotesize Timing and transaction size of the \chainspace implementation of the privacy-preserving e-petition smart contract (described in \Cref{petition}), measured over 10,000 runs. The transactions are independent of the number of authorities. The notation [g] denotes the execution the procedure and [c] denotes the execution of the checker.}
\label{tab:petition}
\end{table}

\subsection{\chainspace Implementation}

We evaluate the \sysname smart contract library implemented in \chainspace, as well as the the coin tumbler (Section~\ref{tumbler}) and the privacy-preserving e-petition (Section~\ref{petition}) applications that use this library.
As expected, \Cref{tab:library} shows that the most time consuming procedures are the checker of \textsf{Create} and the checker of \textsf{Verify}; i.e., they call the \textsf{VerifyCred} primitives which takes about 10 ms (see \Cref{tab:timing}). \Cref{tab:library} is computed assuming two authorities; the transaction size of \textsf{Issue} increases by about 132 bytes (\ie the size of the credentials) for each extra authority\footnote{The \textsf{Request} and \textsf{Issue} procedures are only needed in the case of on-chain issuance (see \Cref{smart_contract_library}).} while the other transactions are independent of the number of authorities.

Similarly, the most time consuming procedure of the coin tumbler (\Cref{tab:tumbler}) application and of the privacy-preserving e-petition (\Cref{tab:petition}) are the checker of \textsf{InitTumbler}  and the checker of \textsf{SignPetition}, respectively; these two checkers call the \textsf{BlindVerify} primitive involving pairing checks. The \textsf{Pay} procedure of the coin tumbler presents the highest transaction size as it is composed of two distinct transactions: a coin transfer transaction and a \textsf{Request} transaction from the \sysname contract library. However, they are all practical, and they all run in a few milliseconds. These transactions are independent of the number of authorities as issuance is either handled off-chain or by the \sysname smart contract library.

\subsection{\ethereum Implementation}

We evaluate the \sysname Ethereum smart contract library described in \Cref{ethereum_smart_contract_library} using the Go implementation of Ethereum on an Intel Core i5 laptop with 12GB of RAM running Ubuntu 17.10. \Cref{tab:ethlibrary} shows the execution times and gas costs for different procedures in the smart contract. The execution times for \algorithm{Create} and \algorithm{Verify} are higher than the execution times for the \chainspace version (\Cref{tab:library}) of the library, due to the different implementations. The arithmetic underlying \sysname in \chainspace is performed through Python naively binding to C libraries, while in \ethereum arithmetic is defined in solidity and executed by the EVM. 

\begin{table}[t]
\begin{center}
    \begin{tabular}{lccc}
        \toprule
        \multicolumn{4}{l}{\sysname Ethereum smart contract library}\\
        \small\bf Operation & \small\bf \boldmath $\mu$ [ms] & \small\bf \boldmath$\sqrt{\sigma^2}$ [ms] & \bf gas\\
        \midrule
        \textsf{Create} & 27.45 & $\pm$ 3.054 & $\sim23,000$\\ 
        \textsf{Verify} & 120.17 & $\pm$ 25.133 & $\sim2,150,000$\\ 
        \bottomrule
    \end{tabular}
\end{center}
\caption{\footnotesize Timing and gas cost of the \ethereum implementation of the \sysname smart contract library described in \Cref{ethereum_smart_contract_library}. Measured over 100 runs, for one public attribute. The transactions are independent of the number of authorities.}
\label{tab:ethlibrary}
\end{table}

We also observe that the \algorithm{Verify} function has a significantly higher gas cost than \algorithm{Create}. This is mostly due to the implementation of elliptic curve multiplication as a native Ethereum 
smart contract---the elliptic curve multiplication alone costs around $1,700,000$ gas, accounting for the vast majority of the gas cost, whereas the pairing operation using the pre-compiled contract costs only 260,000 gas. The actual fiat USD costs corresponding to those gas costs, fluctuate wildly depending on the price of Ether---Ethereum's internal value token---the load on the network, and how long the user wants to wait for the transaction to be mined into a block. As of February 7th 2018, for a transaction to be confirmed within 6 minutes, the transaction fee for \algorithm{Verify} is \$1.74, whereas within 45 seconds, the transaction fee is \$43.5.\footnote{\url{https://ethgasstation.info/}}

The bottleneck of our \ethereum implementation is the high-level arithmetic in $\mathbb{G}_2$. However, \ethereum provides a pre-compiled contract for arithmetic operations in $\mathbb{G}_1$. We could re-write our cryptographic primitives by swapping all the operations in $\mathbb{G}_1$ and $\mathbb{G}_2$, at the cost of relying on the SXDH assumption~\cite{ramanna2016efficient} (which is stronger than the standard XDH assumption that we are currently using).

%% file: sections/related.tex
\section{Comparison with Related Works} \label{related}

\newcolumntype{L}{l}
\newcolumntype{C}{c}
\definecolor{verylightgray}{gray}{0.9}
\begin{table*}[t]
\begin{center}
    \begin{tabular}{L C C C C C}
        \toprule
        \small\bf Scheme & \small\bf Blindness & \small\bf Unlinkable & \small\bf Aggregable  & \small\bf Threshold  & \small\bf Size \\
        \midrule
        
        \textbf{\cite{waters}} Waters Signature & \no & \no & \L & \no & 2 Elements\\
        \textbf{\cite{lossw}} LOSSW Signature & \no & \no & \M & \no & 2 Elements\\
        \textbf{\cite{bgls}} BGLS Signature & \yes & \no & \H & \yes & 1 Element\\
        \textbf{\cite{cl}} CL Signature & \yes & \yes & \M & \no & $O(q)$ Elements\\ 
        \textbf{\cite{idemix}} Idemix & \yes & \yes & \L & \no & $O(q)$ Elements \\
        \textbf{\cite{uprove}} U-Prove & \yes & \yes & \L & \no & $O(v)$ Elements\\ 
         \textbf{\cite{acl}} ACL & \yes & \yes & \L & \no & $O(v)$ Elements\\ 
        \textbf{\cite{pointcheval}} Pointcheval and Sanders & \yes & \yes & \M & \no & 2 Elements\\ 
        \textbf{\cite{dac}} Garman~\etal & \yes & \yes & - & \no & 2 Elements\\ 
        \rowcolor{verylightgray} \textbf{[\Cref{construction}]} \sysname & \yes & \yes & \H & \yes & 2 Elements\\ 

        \bottomrule
    \end{tabular}
\end{center}
\caption{\footnotesize Comparison of \sysname with other relevant cryptographic constructions. The aggregability of the signature scheme reads as follows;  \L: not aggregable,  \M: sequentially aggregable,  \H: aggregable. The signature size is measured asymptotically or in terms of the number of group elements it is made of (for constant-size credentials); $q$ indicates the number of attributes embedded in the credentials and $v$ the number of times the credential may be shown unlinkably.}
\label{tab:related_works}
\end{table*}

We compare the \sysname cryptographic constructions and system with related work in \Cref{tab:related_works}, along the dimensions of key properties offered by \sysname---blindness, unlinkability, aggregability (i.e., whether multiple authorities are involved in issuing the credential), threshold aggregation (i.e., whether a credential can be aggregated using signatures issued by a subset of authorities), and signature size (see Sections~\ref{architecture}~and~\ref{construction}).

\paragraph{Short and aggregable signatures} The Waters signature scheme~\cite{waters} provides the bone structure of our primitive, and introduces a clever solution to aggregate multiple attributes into short signatures. 
However, the original Waters signatures do not allow blind issuance or unlinkability, and are not aggregable as they have not been built for use in a multi-authority setting. 
Lu~\etal scheme, commonly known as LOSSW signature scheme~\cite{lossw}, is also based on Waters scheme and comes with the improvement of being sequentially aggregable. 
In a sequential aggregate signature scheme, the aggregate signature is built in turns by each signing authority; this requires the authorities to communicate with each other resulting in increased latency and cost. The BGLS signature~\cite{bgls} scheme is built upon BLS signatures and is remarkable because of its short signature size---signatures are composed of only one group element. The BGLS scheme has a number of desirable properties as it is aggregable without needing coordination between the signing authorities, and can be extended to work in a threshold setting~\cite{boldyreva2002efficient}. Moreover, Boneh~\etal show how to build verifiably encrypted signatures~\cite{bgls} which is close to our requirements, but not suitable for anonymous credentials.

\paragraph{Anonymous credentials} CL Signatures~\cite{cl,lee2013aggregating} and Idemix~\cite{idemix} are amongst the most well-known building blocks that inspired applications going from direct anonymous attestations~\cite{chen2010design, bernhard2013anonymous} to electronic cash~\cite{canard2015divisible}. They provide blind issuance and unlikability through randomization; but come with significant computational overhead and credentials are not short as their size grows linearly with the number of signed attributes, and are not aggregable. U-Prove~\cite{uprove} and Anonymous Credentials Light (ACL)~\cite{acl} are computationally efficient credentials that can be used once unlinkably; therefore the size of the credentials is linear in the number of unlinkable uses. Pointcheval and Sanders~\cite{pointcheval} present a construction which is the missing piece of the BGLS signature scheme; it achieves blindness by allowing signatures on committed values and unlinkability through signature randomization. However, it only supports sequential aggregation and does not provide threshold aggregation. For anonymous credentials in a setting where the signing authorities are also verifiers (i.e., without public verifiability), Chase~\etal~\cite{amac} develop an efficient protocol. Its `GGM' variant has a similar structure to \sysname, but forgoes the pairing operation by using message authentication codes (MACs). None of the above schemes support threshold issuance.

While the scheme of Garman~\etal~\cite{dac} does not specifically focus on threshold issuance of credentials or on general purpose credentials, it provides the ability to issue credentials without central issuers supporting private attributes, blind issuance, and unlinkable multi-show selective disclosure. To obtain a credential, users build a vector commitment to their secret key and a set of attributes; and append it to a ledger along with a pseudonym built from the same secret key, and a zk-proof asserting the correctness of the vector commitment and of the pseudonym. To show a credential under a different pseudonym, users scan the ledger for all credentials and build a RSA accumulator; they provide a zk-proof that they know a credential embedded in the accumulator. Similarly to Zerocoin~\cite{zerocoin}, showing credentials requires an expensive double  discrete-logarithm proof (about 50KB~\cite{dac}); and the security of the credentials scheme relies on the security of the ledger. \sysname addresses the two open questions left as future work by Garman~\etal~\cite{dac}; \first the security of \sysname credentials do not depend on the security of a transaction ledger as they are general purpose credentials, and \second \sysname enjoys short and efficient proofs as it builds from blind signatures and does not require cryptographic accumulators.

\paragraph{Short and threshold issuance anonymous credentials} \sysname extends these previous works by presenting a short, aggregable, and randomizable credential scheme; allowing threshold and blind issuance, and a multi-authority anonymous credentials scheme. \sysname primitives do not require  sequential aggregation, that is the aggregate operation does not have to be performed by each signer in turn. Any independent party can aggregate any threshold number of partial signatures into a single aggregate credential, and verify its validity. 





%% file: sections/limitations.tex
\section{Limitations} \label{limitations}

\sysname has a number of limitations that are beyond the scope of this work, and deferred to future work. 

Adding and removing authorities implies to re-run the key generation algorithm---this limitation is inherited from the underlying Shamir's secret sharing protocol~\cite{shamir1979share} and can be mitigated using techniques coming from proactive secret sharing introduced by Herzberg~\etal~\cite{herzberg1995proactive}. However, credentials issued by authorities with different key sets are distinguishable, and therefore frequent key rotation reduces the privacy provided. 

As any threshold system, \sysname is vulnerable if more than the threshold number of authorities are malicious; colluding authorities could create coins to steal all the coins in the buffer of the coin tumbler application (\Cref{tumbler}), create fake identities or censor legitimate users of the electronic petition application (\Cref{petition}), and defeat the censorship resistance of our proxy distribution application (\Cref{proxy}). Note that users' privacy is still guaranteed under colluding authorities, or an eventual compromise of their keys.

Implementing the \sysname smart contract library on \ethereum is expensive (\Cref{tab:ethlibrary}) as \ethereum does not provide pre-compiled contracts for elliptic curve arithmetic in $\mathbb{G}_2$; re-writing our cryptographic primitives by swapping all the operations in $\mathbb{G}_1$ and $\mathbb{G}_2$ would dramatically reduce the gas cost, at the cost of relying on the SXDH assumption~\cite{ramanna2016efficient}.

%% file: sections/conclusion.tex
\section{Conclusion} \label{conclusion}



Existing selective credential disclosure schemes do not provide the full set of desired properties, particularly when it comes to efficiency and issuing general purpose selective disclosure credentials without sacrificing desirable distributed trust assumptions. This limits their applicability in distributed settings such as distributed ledgers, and prevents security engineers from implementing separation of duty policies that are effective in preserving integrity. In this paper, we present \sysname---a novel scheme that supports distributed threshold issuance, public and private attributes, re-randomization, and multiple unlinkable selective attribute revelations. 
We provide an overview of the \sysname system, and the cryptographic primitives underlying \sysname;
an implementation and evaluation of \sysname as a smart contract library in \chainspace and \ethereum, a sharded and a permissionless blockchain respectively; and three diverse and important application to anonymous payments, petitions and censorship resistance. \sysname fills an important gap in the literature and enables general purpose selective disclosure credentials---an important privacy enhancing technology---to be efficiently used in settings with no natural single trusted third party to issue them, and to interoperate with modern transparent computation platforms.

%% file: acknowledgements.tex
\section*{Acknowledgements} George Danezis, Shehar Bano and Alberto Sonnino are supported in part by EPSRC Grant EP/N028104/1 and the EU H2020 DECODE project under grant agreement number 732546 as well as \texttt{chainspace.io}. Mustafa Al-Bassam is supported by The Alan Turing Institute. Sarah Meiklejohn is supported in part by EPSRC Grant EP/N028104/1. We thank Jonathan Bootle, Andrea Cerulli, Natalie Eskinazi, and Moxie Marlinspike for helpful suggestions on early manuscripts. We extend our thanks the anonymous reviewers for their valuable advice, and particularly to Christina Garman for kindly accepting to shepherd the paper and her many insightful comments and suggestions.

%% file: appendices/security.tex
\section{Sketch of Security Proofs} \label{security_proofs}
This appendix sketches the security proofs of the cryptographic construction described in \Cref{construction}.

\paragraph{Unforgeability} There are two possible ways for an adversary to forge a proof of a credential: \first an adversary without a valid credential nevertheless manages to form a proof such that \algorithm{VerifyCred} passes; and \second, an adversary that has successfully interacted with fewer than $t$ authorities generates a valid consolidated credential (of which they then honestly prove possession using \algorithm{ProveCred}).

Unforgeability in scenario \first is ensured by the soundness property of the zero-knowledge proof.  For scenario \second, running \algorithm{AggCred} involves performing Lagrange interpolation.  If an adversary has fewer than $t$ partial credentials, then they have fewer than $t$ points, which makes the resulting polynomial (of degree $t-1$) undetermined and information-theoretically impossible to compute.  The only option available to the adversary is thus to forge the remaining credentials directly.  This violates the unforgeability of the underlying blind signature scheme, which was proved secure by Pointcheval and Sanders~\cite{pointcheval} under the LRSW assumption~\cite{lysyanskaya1999pseudonym}.

\paragraph{Blindness} Blindness follows directly from the blindness of the signature scheme used during \algorithm{IssueCred}, which was largely proved secure by Pointcheval and Sanders~\cite{pointcheval} under the XDH assumption~\cite{bls}.  There are only two differences between their protocol and ours.

First, the \sysname authorities generate the credentials from a group element $h=\hashtopoint(c_m)$ instead of from $g_1^{\tilde{r}}$ for random $\tilde{r} \in \mathbb{F}_p$.  The hiding property of the commitment $c_m$, however, ensures that $\hashtopoint(c_m)$ does not reveal any information about $m$. Second, Pointcheval and Sanders use a commitment to the attributes as input to \algorithm{BlindSign} (see \Cref{pointcheval_recall}), whereas \sysname uses an encryption instead.  The IND-CPA property, however, of the encryption scheme ensures that the ciphertext also reveals no information about $m$.

Concretely, \sysname uses Pedersen Commitments~\cite{pedersen} for the commitment scheme, which is secure under the discret logarithm assumption.  It uses \elgamal for the encryption scheme in $\mathbb{G}_1$, which is secure assuming DDH.  Finally, it relies on the blindness of the Pointcheval and Sanders signature, which is secure assuming XDH~\cite{bls}.  As XDH implies both of the previous two assumptions, our entire blindness argument is implied by XDH.


\paragraph{Unlinkability / Zero-knowledge} Unlinkability and zero-knowledge are guaranteed under the XDH assumption~\cite{bls}. The zero-knowledge property of the underlying proof system ensures that \algorithm{ProveCred} does not on its own reveal anything more than the validity of the statement $\phi'$, which may include public attributes (see \Cref{multi_message_scheme} and Appendix \ref{multi_attribute_credentials}). The fact that credentials are re-randomized at the start of \algorithm{ProveCred} in turn ensures unlinkability, both between different executions of \algorithm{ProveCred} and between an execution of \algorithm{ProveCred} and of \algorithm{IssueCred}.


%% file: appendices/multi_attribute_credentials.tex
\section{Multi-Attributes Credentials} \label{multi_attribute_credentials}
We present the cryptographic primitives behind the multi-attribute \sysname threshold issuance credential scheme described in \Cref{multi_message_scheme}. As in \Cref{threshold_credentials_scheme}, we describe below a key generation algorithm \algorithm{TTPKeyGen} as executed by a trusted third party; this protocol can however be execute in a distributed way as illustrated by Kate~\etal~\cite{cryptoeprint:2012:377}.

\begin{description}[leftmargin=1em, labelindent=0em]
\setlength\itemsep{0.5em}
\item[\definition{Setup($1^\lambda,q$)}{$params$}] Choose a bilinear group $(\mathbb{G}_1,\mathbb{G}_2,\mathbb{G}_T)$ with order $p$, where $p$ is an $\lambda$-bit prime number. Let $g_1, h_1, \dots, h_{q}$ be generators of $\mathbb{G}_1$, and $g_2$ a generator of $\mathbb{G}_2$. The system parameters are $params=(\mathbb{G}_1, \mathbb{G}_2, \mathbb{G}_T, p, g_1, g_2, h_1, \dots, h_{q} )$.

\item[\definition{TTPKeyGen($params, t, n, q$)}{$sk,vk$}] Choose $(q+1)$ polynomials of degree $(t-1)$ with coefficients in $\mathbb{F}_p$, noted $(v, w_1, \dots, w_{q})$,  and set:
\begin{equation*}
(x,y_1, \dots y_{q}) = (v(0),w_1(0), \dots, w_{q}(0))
\end{equation*}
Issue a secret key $sk_i$ to each authority $i \in [1, \dots, n]$ as below:
\begin{equation*}
sk_i = (x_i,y_{i,1}, \dots, y_{i,q}) = (v(i), w_{1}(i),\dots, w_{q}(i))
\end{equation*}
and publish their verification key $vk_i$ computed as follows:
\begin{equation*}
vk_i = (g_2,\alpha_i,\beta_{i,1}, \dots, \beta_{i,q)}) = (g_2,g_2^{x_i},g_2^{y_{i,1}}, \dots, g_2^{y_{i,1q}})
\end{equation*}

\item[\definition{IssueCred($m_1,\dots,m_{q}, \phi$)}{$\sigma$}] Credentials issuance is composed of three algorithms:
\begin{description}[leftmargin=1em, labelindent=0em]
\setlength\itemsep{0.5em}
\item \definition{PrepareBlindSign($m_1, \dots, m_q,\phi$)}{$d,\Lambda,\phi$} The users generate an \elgamal key-pair $(d, \gamma=g_1^{d})$; pick a random $o\in\mathbb{F}_p$ compute the commitment $c_m$ and the group element $h\in\mathbb{G}_1$ as follows:
\begin{equation}\nonumber
c_m = g_1^o \prod_{j=1}^{q} h_j^{m_j} \qquad{\rm and}\qquad h = \hashtopoint(c_m)
\end{equation} 
Pick at random $(k_1,\dots,k_{q}) \in \mathbb{F}_p^{q}$ and compute an \elgamal encryption of each $m_j$ for $\forall j \in [1, \dots, q]$  as below:
\begin{equation}\nonumber
c_j = Enc(h^{m_j})=(g_1^{k_j},\gamma^{k_j} h^{m_j})
\end{equation}
Output ($d,\Lambda=(\gamma, c_m, c_j, \pi_{s}),\phi$) $\forall j \in [1, \dots, q]$, where $\pi_{s}$ is defined by:
\begin{eqnarray}\nonumber
\pi_{s} &=& {\rm NIZK}\{(d, m_1,\dots,m_q, o, k_1,\dots,k_q): \gamma = g_1^d \\ \nonumber
 &&\land\; c_m = g_1^o \prod_{j=1}^{q} h_j^{m_j}  \;\land\; c_j = (g_1^{k_j,}\gamma^{k_j} h^{m_j}) \\ \nonumber
 && \land\;  \phi(m_1,\dots,m_{q})=1\} \quad \forall j \in [1, \dots, q] 
 \end{eqnarray}
 
\item \definition{BlindSign($sk, \Lambda, \phi$)}{$\tilde{\sigma}_i$} The authority $i$ parses $\Lambda=(\gamma, c_m, c_j, \pi_{s})$ and $c_j=(a_j,b_j)$ $\forall j \in [1, \dots, q]$, and $sk_i=(x,y_1, \dots, y_{q})$. Recompute $h = \hashtopoint(c_m)$. Verify the proof  $\pi_{s}$ using $\gamma,c_m$ and $\phi$. If the proof is invalid, output $\perp$ and stop the protocol; otherwise output $\tilde{\sigma}_i=(h,\tilde{c})$, where $\tilde{c}$ is defined as below:
\begin{equation}\nonumber
\tilde{c}=\left(\prod_{j=1}^{q}a_j^{y_j}, h^x \prod_{j=1}^{q}b_j^{y_j}\right)
\end{equation}

\item \definition{Unblind($\tilde{\sigma}_i, d$)}{$\sigma_i$} The users parse $\tilde{\sigma}_i=(h, \tilde{c})$ and $\tilde{c}=(\tilde{a},\tilde{b})$; compute $\sigma_i = (h,\tilde{b}(\tilde{a})^{-d})$. Output $\sigma_i$.
 \end{description}

\item[\definition{AggCred($\sigma_1, \dots, \sigma_t$)}{$\sigma$}] Parse each $\sigma_i$ as $(h,s_i)$ for $i \in [1, \dots, t]$. Output $(h,\prod^t_{i=1} s_i^{l_i})$, where:
\begin{equation}\nonumber
l_i = \left[\prod^t_{i=1, j\neq i} (0-j)\right] \left[\prod^t_{i=1,s j\neq i} (i-j)\right]^{-1} \;{\rm mod}\; p
\end{equation}

\item[\definition{ProveCred($vk, m_1, \dots, m_{q}, \sigma, \phi'$)}{$\sigma',\Theta,\phi'$}] Parse $\sigma=(h,s)$ and $vk=(g_2,\alpha,\beta_{1}, \dots, \beta_{q})$. Pick at random $r',r\in\mathbb{F}_q^2$; set $\sigma'=(h',s')=(h^{r'},s^{r'})$, and build $\kappa$ and $\nu$ as below:
\begin{equation}\nonumber
\kappa =  \alpha\prod_{j=1}^{q}{\beta_j^{m_j}} g_2^r \qquad{\rm and}\qquad \nu= \left(h'\right)^r
 \end{equation}
 Output $(\Theta=(\kappa, \nu, \sigma', \pi_v),\phi')$, where $\pi_v$ is:
\begin{eqnarray}\nonumber
\pi_{v} &=& {\rm NIZK}\{(m_1,\dots,m_q,r): \kappa= \alpha\prod_{j=1}^{q}{\beta_j^{m_j}} g_2^r \\ \nonumber
 && \ \land \ \nu=\left(h'\right)^r \ \land \  \phi(m_1, \dots, m_{q})=1\} 
 \end{eqnarray}

\item[\definition{VerifyCred($vk, \Theta, \phi'$)}{$true/false$}] Parse $\Theta=(\kappa, \nu, \sigma',  \pi_v)$ and $\sigma'=(h',s')$; verify $\pi_v$ using $vk$ and $\phi'$; Output $true$ if the proof verifies, $h'\neq1$ and $e(h',\kappa)=e(s'\nu,g_2)$; otherwise output $false$.
\end{description}

%% file: appendices/eth-tumbler.tex
\section{Ethereum tumbler}\label{eth-tumbler}

We extend the example of the tumbler application described in \Cref{tumbler} to the Ethereum version of the \sysname library, with a few modifications to reduce the gas costs.

Instead of having $v$ (the number of coins) as an attribute, which would increase the number of elliptic curve multipications required to verify the credentials, we allow for a fixed number of instances of \sysname to be setup for different denominations for $v$. The Tumbler has a \algorithm{Deposit} method, where users deposit Ether into the contract, and then send an issuance request to authorities on one private attribute: $addr || s$, where $addr$ is the destination address of the merchant, and $s$ is a randomly generated sequence number~(1). It is necessary for $addr$ to be a part of the attribute because once the attribute is revealed, the credential can be spent by anyone with knowledge of the attribute (including any peers monitoring the blockchain for transactions), therefore the credential must be bounded to a specific recipient address before it is revealed. This issuance request is signed by the Ethereum address that deposited the Ether into the smart contract, as proof that the request is associated with a valid deposit, and sent to the authorities~(2). As $addr$ and $s$ will be both revealed at the same time when withdrawing the token, we concatenate these in one attribute to save on elliptic curve operations. Users aggregate the credentials issued by the authorities~(3). The resulting token can then be passed to the \algorithm{Withdraw} function, where the withdrawer reveals $addr$ and $s$~(4). As usual, the contract maintains a map of $s$ values associated with tokens that have already been withdrawn to prevent double-spending. After checking that the token's credentials verifies and that it has not already been spent, the contract sends $v$ to the Ethereum destination address $addr$~(5).